\documentclass[%
preprint,
nofootinbib,
 amsmath,amssymb,
 aps,
]{revtex4-1}

\usepackage{rotating}
\usepackage{graphicx}
\usepackage{dcolumn}
\usepackage{bm}
\usepackage{color}
\usepackage[normalem]{ulem}
\usepackage[makeroom]{cancel}
\usepackage{extarrows}

\def\slashP{P\!\!\! \slash}

\newcommand{\cref}[1]{Chapter~\ref{ch:.#1}}

\newcommand{\beq}{\begin{equation}} 
\newcommand{\eeq}{\end{equation}} 
\newcommand{\ba}{\begin{array}}  
\newcommand{\ea}{\end{array}} 
\newcommand{\bea}{\begin{eqnarray}}  
\newcommand{\eea}{\end{eqnarray} }  
\newcommand{\be}{\begin{eqnarray}}  
\newcommand{\ee}{\end{eqnarray} }  
\newcommand{\bal}{\begin{align}}
\newcommand{\eal}{\end{align}}   
\newcommand{\bi}{\begin{itemize}}  
\newcommand{\ei}{\end{itemize}}  
\newcommand{\ben}{\begin{enumerate}}  
\newcommand{\een}{\end{enumerate}}  
\newcommand{\bc}{\begin{center}}
\newcommand{\ec}{\end{center}} 
\newcommand{\bt}{\begin{table}}
\newcommand{\et}{\end{table}}  
\newcommand{\btb}{\begin{tabular}}
\newcommand{\etb}{\end{tabular}}

\newcommand{\eL}{\epsilon_L}
\newcommand{\eR}{\epsilon_R}
\newcommand{\eS}{\epsilon_S}
\newcommand{\eP}{\epsilon_P}
\newcommand{\eT}{\epsilon_T}
\newcommand{\ket}[1]{| #1 \rangle}
\newcommand{\bra}[1]{\langle #1 |}

\begin{document}

\preprint{INT-PUB-16-014}
\preprint{MITP/16-047}

\title{Global Effective-Field-Theory analysis of New-Physics effects in (semi)leptonic kaon  decays}

\author{Mart\'{i}n Gonz\'{a}lez-Alonso}%
\affiliation{%
IPN de Lyon/CNRS, Universite Lyon 1, Villeurbanne, France
}
\author{Jorge Martin Camalich}
\affiliation{PRISMA Cluster of Excellence Institut f\"ur Kernphysik, 
Johannes Gutenberg-Universit\"at Mainz, 55128 Mainz, Germany}

\begin{abstract}

We analyze the decays $K\to\pi\ell\nu$ and $P\to\ell\nu$ ($P=K,\pi$, $\ell=e,\,\mu$) using a low-energy Effective-Field-Theory approach to 
parametrize New Physics and study the complementarity with baryon $\beta$ decays. We then provide a road map for a global 
analysis of the experimental data, with all the Wilson coefficients simultaneously, and perform a fit leading to numerical bounds for 
them and for $V_{us}$. A prominent result of our analysis is a reinterpretation of the well-known $V_{ud}-V_{us}$ diagram
as a strong constraint on new physics. Finally, we reinterpret our bounds in terms 
of the $SU(2)_L\times~U(1)_Y$-invariant operators, provide bounds to the corresponding Wilson coefficients at the TeV scale 
and compare our results with collider searches at the LHC.  

\end{abstract}

\pacs{Valid PACS appear here}
\maketitle


\section{Introduction}
\label{sec:Intro}

The $K\to\pi\ell\nu$ ($K_{\ell3}$) and $P\to \ell\nu$ ($P_{\ell2}$) decays, where $P=\pi,K$ and $\ell=e,\,\mu$, boast one of the most precise data bases
in hadronic weak decays~\cite{Antonelli:2008jg,Antonelli:2010yf,Cirigliano:2011ny,Agashe:2014kda}. The hadronic form factors necessary to describe these processes are flagship quantities for
lattice QCD (LQCD) and the theoretical accuracy at which they are calculated is now below the percent level (relative uncertainty)~\cite{Aoki:2013ldr,Rosner:2015wva}. 
Much work has also been done in Chiral Perturbation Theory (ChPT) and using dispersive methods to understand analytically low-energy theorems and small 
contributions to the decay rates such as isospin breaking and the electromagnetic radiative corrections~\cite{Gasser:1984gg,Gasser:1984ux,Cirigliano:2001mk,
Bijnens:2003uy,Jamin:2001zq,Ananthanarayan:2004qk,Jamin:2004re,Cirigliano:2005xn,DescotesGenon:2005pw,Cirigliano:2007xi,Cirigliano:2007ga,Kastner:2008ch,
Cirigliano:2008wn,Bernard:2009zm,Cirigliano:2011tm}. This makes 
$P_{\ell2}$ and $K_{\ell3}$ ideal flavor benchmarks to test the Cabibbo-Kobayashi-Maskawa (CKM) structure in the Standard Model (SM) and to search for new physics (NP). 
In fact, a systematic search of NP effects in (semi)leptonic kaon decays is particularly interesting at the moment
as several anomalies have been recently reported in flavor observables such as $B$ decay 
rates~\cite{Lees:2012xj,Lees:2013uzd,Aaij:2014ora,Aaij:2015yra,Huschle:2015rga,Aaij:2015oid,Abdesselam:2016cgx,Abdesselam:2016llu}
or $K\bar K$ mixing~\cite{Buras:2015yba}.   

An optimal tool to perform such analysis is that of Effective Field Theory (EFT), which allows us to test the SM in a model-independent way. 
In fact, in addition to studies within specific NP scenarios~\cite{Masiero:2005wr,Bernard:2006gy,Bernard:2007cf,Campbell:2008um,Antonelli:2008jg,Antonelli:2010yf,Carpentier:2010ue,Jung:2010ik,Bauman:2012fx}, the EFT language has been introduced~\cite{Antonelli:2008jg,Cirigliano:2009wk,Antonelli:2010yf}. 
%
However, an EFT approach has not been used yet for global studies of the $s\to u$ transitions beyond the $U(3)^5$-symmetric limit~\footnote{
$U(3)^5$ refers to the flavor symmetry of the SM gauge Lagrangian, i.e. the freedom to perform $U(3)$ rotations in family space for 
each of the five fermionic gauge multiplets: $Q_L = (u_L, d_L)$, $u_R$, $d_R$, $L_L = (\nu_L, e_L)$, $e_R$.} where the only NP probe is the
CKM unitarity test, given by precise determinations of $|V_{us}|$ and $|V_{ud}|$~\cite{Cirigliano:2009wk}. Notice the difference with 
the $d\to u$ decays, where global EFT fits have been performed by various groups~\cite{Severijns:2006dr,Wauters:2013loa,Pattie:2013gka}.

In this paper we amend these limitations, giving the natural next step in the analysis of (semi)leptonic kaon decays:
\begin{itemize}
\item We do not assume any flavor symmetry, generalizing in this way the phenomenological EFT analysis performed in the $U(3)^5$-symmetric limit in Ref.~\cite{Cirigliano:2009wk}. 
\item We keep all operators at the same time. Notice that non-trivial correlations are possible, not only between NP Wilson coefficients (WC) 
but also involving QCD parameters that are extracted phenomenologically. This generalizes previous works~\cite{Bernard:2006gy,Antonelli:2008jg,Antonelli:2010yf}, 
which are covered by our study as specific cases, as we will explicitly show.
\item We investigate the complementarity with nuclear, neutron and hyperon $\beta$ decays, which are driven by the same underlying $D\to u\ell\nu$ transition ($D=s,\,d$).
\item We provide numerical bounds for the WC. They are to be confirmed by the experimental collaborations taking into
account certain correlations not publicly available.
\item We match with the so-called SMEFT, i.e. the EFT of the SM at the electroweak (EW) scale, with a linear realization of 
the electroweak symmetry breaking~\cite{Buchmuller:1985jz}. This makes possible to study the interplay with high-energy searches, 
as it was done in Ref.~\cite{Cirigliano:2009wk} in the limit of the flavor symmetry $U(3)^5$. Notice that these flavor-physics studies are fairly clean 
probes of a small number of WC (compared with searches in colliders). Thus, an interesting degree of complementarity is expected.  
\end{itemize}

Let us stress that our analysis includes the SM limit as a specific case. In fact our output are not only the bounds on the 
various WC, but also the $V_{us}$ and $V_{ud}$ CKM matrix elements, and includes various QCD form factors parameters. In the SM 
limit we recover the most precise of them~\cite{Antonelli:2010yf,Moulson:2014cra}, with small improvements due to the inclusion
of the individual rate of $K_{\mu2}$ as a separate input and the Callan-Treiman theorem.

The outline of the paper is the following. In Sec.~\ref{sec:EFT} we briefly introduce the EFT framework, which we use in 
Sec.~\ref{sec:decayobs} to analyze the channels $K\to\pi\ell\nu$, $P\to\ell\nu$ ($P=K,\pi$) and baryon $\beta$ decays. This 
section contains all the relevant formulas, expressing our observables in terms of the parameters of our fit. Sec.~\ref{sec:strategy}
describes how to analyze experimental data with all WC present simultaneously. Sec.~\ref{sec:Pheno} describe the numerical aspects of our
analysis and the results of our fit. Then, Sec.~\ref{sec:SMEFT} contains the running of our bounds to the EW scale, the translation to the SMEFT 
WC and a brief comparison with LHC searches. Finally, in Sec.~\ref{sec:Conclusions} we conclude.

\section{The low-energy effective Lagrangian}
\label{sec:EFT}

The low-scale  $O(1 \ {\rm GeV})$ effective Lagrangian for $D \to u$ transitions ($D=s,\,d$) is~\cite{Cirigliano:2009wk}:
\begin{eqnarray}
{\cal L}_{\rm eff} 
&=&
- \frac{G_F^{(0)} V_{uD}}{\sqrt{2}}  \,\sum_{\ell=e,\mu}
\Bigg[
\Big(1 + \epsilon_L^{D \ell}  \Big) \bar{\ell}  \gamma_\mu  (1 - \gamma_5)   \nu_{\ell} \cdot \bar{u}   \gamma^\mu (1 - \gamma_5 ) D
+  \eR^{D\ell}  \   \bar{\ell} \gamma_\mu (1 - \gamma_5)  \nu_\ell    \ \bar{u} \gamma^\mu (1 + \gamma_5) D\nonumber\\
&&+~ \bar{\ell}  (1 - \gamma_5) \nu_{\ell} \cdot \bar{u}  \Big[  \epsilon_S^{D\ell}  -   \eP^{D\ell} \gamma_5 \Big]  D
+ \epsilon_T^{D\ell}    \,   \bar{\ell}   \sigma_{\mu \nu} (1 - \gamma_5) \nu_{\ell}    \cdot  \bar{u}   \sigma^{\mu \nu} (1 - \gamma_5) D
\Bigg]+{\rm h.c.}, 
\label{eq:leff1} 
\end{eqnarray}
where we use $\sigma^{\mu \nu} = i\,[\gamma^\mu, \gamma^\nu]/2$ and $G_F^{(0)}\equiv \sqrt{2}g^2/(8 M_W^2)$
is the tree-level definition of the Fermi constant. The latter is obtained from muon decay, which can be also 
affected by NP effects entering through the normalization of the $\mu\to e\nu_\mu\bar\nu_e$ effective vertex~\cite{Cirigliano:2009wk}:
\begin{equation}
 G_\mu=G_F^{(0)}(1+\frac{\delta G_F}{G_F}).\label{eq:GF}
\end{equation}

In the derivation of the effective Lagrangian in Eq.~(\ref{eq:leff1}) we have assumed that potential right-handed neutrino fields 
(sterile with respect to the SM gauge group) are heavy compared to the low-energy scale.~\footnote{Let us notice that the inclusion 
of operators with right-handed neutrinos is not expected to affect our results, as they do not interfere with the 
SM amplitude and thus contribute at $\mathcal{O}(\epsilon_i^2)$ to the observables.} We focus on $CP$-even observables,
and therefore only the real parts of the WC will interfere with the SM. For the sake of brevity we will write simply $\epsilon_i$
instead of $\rm{Re}(\epsilon_i)$ hereafter.

The $\epsilon_i^{D\ell}$ coefficients carry a $\sim v^2/\Lambda^2$ dependence on the NP scale $\Lambda$ and in the SM 
they vanish leaving the $V-A$ structure generated by the exchange of a $W$ boson.
If the NP is coming from dynamics at $\Lambda \gg v$ and electro-weak symmetry breaking is linearly realized,
then one can use an $SU(2)_L\times U(1)_Y$ invariant effective theory~\cite{Buchmuller:1985jz,Cirigliano:2009wk,Alonso:2014csa,Aebischer:2015fzz}. 
In this case~\cite{Bernard:2006gy,Cirigliano:2009wk,Alonso:2015sja}:
\begin{equation}
\epsilon_R^{De}=\epsilon_R^{D\mu}+\mathcal{O}(v^4/\Lambda^4)\equiv \epsilon_R^{D},\label{eq:RHCuniversal} 
\end{equation}
so that, up to a subleading corrections in the EFT expansion, a NP effect involving a right-handed current necessarily involves
a Higgs-current fermion-current  operator~\cite{Buchmuller:1985jz} and its contribution must be lepton universal. 

Taking into account the points above, and working to linear order in the NP couplings,
we can re-express the Lagrangian~(\ref{eq:leff1}) as~\cite{Bhattacharya:2011qm}:
\begin{eqnarray}
{\cal L}_{\rm eff} 
&=&
- \frac{G_F \tilde V_{uD}^\ell}{\sqrt{2}} \,
\Bigg[
\bar{\ell}  \gamma_\mu  (1 - \gamma_5)   \nu_{\ell} 
\cdot \bar{u}   \Big[ \gamma^\mu \ - \  \big(1 -2  \epsilon_R^{D}  \big)  \gamma^\mu \gamma_5 \Big] D \label{eq:leff2} \\
&+& \bar{\ell}  (1 - \gamma_5) \nu_{\ell}
\cdot \bar{u}  \Big[  \epsilon_S^{D\ell}  -   \eP^{D\ell} \gamma_5 \Big]  D
+ \epsilon_T^{D\ell}     \,   \bar{\ell}   \sigma_{\mu \nu} (1 - \gamma_5) \nu_{\ell}    \cdot  \bar{u}   \sigma^{\mu \nu} (1 - \gamma_5) D
\Bigg]+\mathcal{O}(\epsilon^2)+{\rm h.c.}, \nonumber
\end{eqnarray}
where
\begin{equation}
\tilde V_{uD}^\ell=\left(1 + \epsilon_L^{D\ell} + \epsilon_R^{D}-\frac{\delta G_F}{G_F}\right)\,V_{uD}~.
\label{eq:VuDNPs}
\end{equation}
In addition to $\tilde V_{ud}^e$ and $\tilde V_{us}^e$, we have a total of 16 (combinations of) WC describing the NP modifications
to the charged-current decays $D\to u\ell\nu$ in the SM. The form of the Lagrangian is convenient as it allows to separate the effects of a combination of current-current operators
affecting the normalization of the rates and which can be only accessed through CKM-unitarity and lepton-universality tests.

\subsection{Renormalization and scale running of the Wilson coefficients}

The WC display renormalization-scale dependence that is to be canceled in the observables by the opposite dependence in the 
quantum corrections to the matrix elements of the decays. For instance in QCD we have, at one loop:
\begin{eqnarray}
\epsilon_{i}^{D\ell}(\mu)=\left(\frac{\alpha_s(\mu)}{\alpha_s(\Lambda)}\right)^{-\frac{\gamma_i}{\beta_0}} \epsilon_{i}^{D\ell}(\Lambda), \label{eq:epsrun}
\end{eqnarray}
where $\alpha_s$ is the strong structure constant, $\beta_0=11-2/3N_f$ is the one-loop QCD $\beta$-function coefficient for $N_f$ 
dynamical quark flavors, and $\gamma_{L,R}=0$, $\gamma_{S,P}=-4$ and $\gamma_{T}=4/3$ are the one-loop coefficients of the corresponding 
anomalous dimensions.\footnote{In our conventions, at one loop we have 
$\frac{d\epsilon_i}{d\log\mu}=\gamma_i \frac{\alpha_s}{2\pi} \epsilon_i$ and $\frac{d\alpha_s}{d\log\mu}=-\beta_0 \frac{\alpha_s^2}{2\pi}$.}

One can also consider the renormalization of the effective operators with respect to electroweak corrections. Although they are
very small, they are important for the accuracy of the SM predictions~\cite{Sirlin:1981ie,Marciano:1993sh},
and they induce mixing among certain NP operators that can have interesting phenomenological consequences~\cite{Voloshin:1992sn,Herczeg:1994ur,Campbell:2003ir,Alonso:2013hga}. 
In our case it is important to take into account the mixing they induce between the (pseudo)scalar and tensor operators, since $\pi_{\ell2}$ and $K_{\ell 2}$
set very strong bounds on the pseudoscalar couplings. Expressing the results obtained in Ref.~\cite{Campbell:2003ir} in terms of the WC $\vec\epsilon(\mu)=(\eS^{D\ell}(\mu),\,\eP^{D\ell}(\mu),\,\eT^{D\ell}(\mu))$ we obtain:
\begin{align}
\frac{d\,\vec\epsilon(\mu)}{d\log\mu}=\frac{\alpha}{2\pi}\gamma_{\rm ew}\,\vec\epsilon(\mu),
\end{align}
where
\begin{align}
\gamma_{\rm ew}=
\left(
\begin{array}{ccc}
 -\frac{3}{2s_w^2}-\frac{113}{36c_w^2} & -\frac{5}{12c_w^2} & {\scriptstyle3}\left(\frac{3}{s_w^2}+\frac{5}{c_w^2}\right) \\
 -\frac{5}{12c_w^2} & -\frac{3}{2s_w^2}-\frac{113}{36c_w^2} &{\scriptstyle3}\left(\frac{3}{s_w^2}+\frac{5}{c_w^2}\right)  \\
 \frac{1}{32}\left(\frac{3}{s_w^2}+\frac{5}{c_w^2}\right)& \frac{1}{32}\left(\frac{3}{s_w^2}+\frac{5}{c_w^2}\right) & 
 -\frac{3}{s_w^2}-\frac{103}{36c_w^2} \\
\end{array}
\label{eq:EWanomalousdim}
\right),
\end{align}
where $\alpha$ is the electromagnetic structure constant, $s_w^2=\sin^2\theta_W$ and $c_w^2=\cos^2\theta_W$.

\section{Decay observables}
\label{sec:decayobs}

In this section we calculate the various observables relevant for our analysis in terms of the low-energy WC $\epsilon_i$. 
All the calculations are performed at tree level and the only loop effect taken into account will be the (log-enhanced) 
running of the Wilson Coefficients described in the previous section. 

Nonetheless, it is interesting to notice that our expressions are actually valid at any loop order if the $\epsilon_i$ 
couplings were defined at the amplitude level for each channel (see Ref.~\cite{Gonzalez-Alonso:2014eva} for a similar description of Higgs decays). 
The matching to the low-energy EFT can then later performed at any desired order. Since here we match at tree level we do not make this distinction.

\subsection{$\pi_{\ell2(\gamma)}$ and $K_{\ell2(\gamma)}$}
\label{sec:Kl2}

The photon-inclusive~\footnote{Depending on the channel it might be more convenient (experimentally) to define this rate as fully photon-inclusive ($P=\pi$ case) 
or to include only the ``internal-bremsstrahlung'' contribution ($P=K$ case).}
decay rates are given by
\begin{align}
\Gamma_{P_{\ell2}(\gamma)} = &\frac{G_F^2|\tilde V_{uD}^\ell|^2\,f_{P^{\pm}}^2}{8\pi}m_{P^\pm}\,m_\ell^2\left(1-\frac{m_\ell^2}{m_{P^\pm}^2}\right)^2
\left(1+ \delta^{P\ell}_{\rm em}\right)
\left(1+\Delta^P_{\ell2}\right)
,~\label{eq:Kl2QCD}
\end{align}
where $D=d,\,s$ for $P=\pi,\,K$ respectively, $f_{P^{\pm}}$ is the QCD semileptonic decay constant of $P^\pm$, $\delta^{P\ell}_{\rm em}$ is the corresponding
electromagnetic correction and $\Delta^P_{\ell2}$ contains the NP correction not absorbed in $\tilde V_{uD}^\ell$.

The electromagnetic corrections are given by~\cite{Marciano:1993sh,Finkemeier:1995gi,Cirigliano:2007xi,Cirigliano:2007ga}
\begin{equation}
1+\delta^{P\ell}_{\rm em} = S_{\rm ew}\,\left\{1+\frac{\alpha}{\pi}\left[F(m_\ell^2/m_P^2)+\frac{3}{2}\log\frac{m_P}{m_\rho}-c_1^P\right]\right\}
+\mathcal{O}(e^2p^4), \label{eq:Kl2rad}
\end{equation}
where $\alpha$ is the structure constant, $S_{\rm ew}=1.0232(3)$~\cite{Marciano:1993sh} encodes universal short distance corrections to the 
semileptonic transitions in the SM at $\mu=m_\rho$ and $F(z)$ describes the leading universal long-distance 
radiative corrections to a point-like meson~\cite{Sirlin:1981ie,Marciano:1993sh}. The constant $c_1^P$ encodes hadronic structure effects that can be 
calculated in Chiral Perturbation Theory~\cite{Cirigliano:2007xi,Cirigliano:2007ga,DescotesGenon:2005pw}. These corrections are
at the $1-3$\% level with an uncertainty quite smaller than the current one of $f_{P}$. 

The NP contribution enters at tree level from the Lagrangian in Eq.~(\ref{eq:leff2}):
\begin{align}
\Delta^P_{\ell2}&=\left| 1-2\eR^D -\frac{m_{P^\pm}^2}{m_\ell\left(m_D+m_u\right)}\eP^{D\ell}\right
|^2-1\nonumber\\
&=-4\eR^D -\frac{2m_{P^\pm}^2}{m_\ell\left(m_D+m_u\right)}\eP^{D\ell}+\mathcal{O}(v^4/\Lambda^4), \label{eq:epsPl2ell}
\end{align}
where in the second line we have linearized in the WC by expanding up to leading order in the EFT 
expansion.
A very important feature of $P_{\ell 2}$ is its high sensitivity to pseudoscalar contributions because they lift the 
chiral suppression of the SM. This appears in Eqs.~(\ref{eq:Kl2QCD},~\ref{eq:epsPl2ell}) in the coefficients of $\eP^{D\ell}$ which 
are multiplied by the inverse of the lepton masses. Besides that, partial conservation of the axial current (PCAC) implies the appearance
of the quark masses also in the denominators, although always in combination with the meson masses squared in the numerator. Note that
the latter combination can be re-expressed using the Gell-Mann-Renner-Oakes equations as $m_P^2/(m_u+m_D)\simeq\, B_0$, which 
is a nonperturbative parameter that is related to the quark condensate and the pion semileptonic decay constant, 
$B_0(\mu=2\,\text{GeV})=-2\langle \bar u u\rangle/f_\pi^2\simeq2.5$ GeV. Hence, the contributions from the pseudoscalar operators to the 
electronic (muonic) mode are enhanced by a factor $\sim5000$ ($\sim25$) with respect to the SM.
This means that the linearization performed in Eq.~(\ref{eq:epsPl2ell}) is only valid for very small values of 
the WC $\epsilon_i$. We will assume this is the case in the initial numerical analysis, and, afterwards, we will 
discuss how the limits are relaxed once we take into account the very narrow region in the parameter space where quadratic 
corrections dominate.

The theoretical uncertainty in the SM prediction of these decays can be minimized by taking convenient ratios among the four
possible (CP-averaged) channels. The lepton-universality ratios $R_P=\Gamma(P_{e2(\gamma)})/\Gamma(P_{\mu2(\gamma)})$ have been
very accurately predicted in the SM because the decay constants $f_{P^\pm}$ cancel exactly in the ratio and the 
radiative corrections are known up to order $\mathcal{O}(e^2p^4)$ because the constant $c_1^P$ disappears from
$\delta^{Pe}_{\rm em}-\delta^{P\mu}_{\rm em}$~\cite{Cirigliano:2007xi,Cirigliano:2007ga}.
The dependence on the NP coefficients of $R_P$ is:
\begin{align}
\frac{R_P}{\left.R_P\right|_{\rm SM}}
&\equiv\frac{|\tilde V^e_{uD}|^2}{|\tilde V^\mu_{uD}|^2}\left(1+\Delta^P_{e2/\mu2}\right)\label{eq:epsDemu}\\
&=\frac{|\tilde V^e_{uD}|^2}{|\tilde V^\mu_{uD}|^2}\left(1 - 2B_0 \left(\frac{\eP^{De}}{m_e}-\frac{\eP^{D\mu}}{m_\mu}\right)\right)+\mathcal{O}(v^4/\Lambda^4)\label{eq:deltaRPexp1}\\
&=1+2\left(\eL^{De}-\eL^{D\mu}\right)-2B_0\left(\frac{\eP^{De}}{m_e}-\frac{\eP^{D\mu}}{m_\mu}\right)+\mathcal{O}(v^4/\Lambda^4).\label{eq:deltaRPexp2}
\end{align}
Note that the dependence on NP right-handed currents completely disappears at this order as a consequence of Eq.~(\ref{eq:RHCuniversal}). It is convenient to define $\Delta_L^D=\eL^{D\mu}-\eL^{De}$, as this combination of WC will appear several times in our analysis.

The ratio $R_{\ell}=\Gamma(K_{\ell2(\gamma)})/\Gamma(\pi_{\ell2(\gamma)})$ is interesting because $f_K/f_\pi$~\cite{Aoki:2013ldr} is
calculated in the lattice more accurately than the decay constants separately and the combination of radiative corrections entering is
independent of the low-energy constants in the hadron-structure functions $c_1^P$ at $\mathcal{O}(e^2\,p^2)$~\cite{Cirigliano:2011tm}.
From this ratio one can obtain
\begin{align}
R_\ell 
\to~ 
& \frac{|\tilde V_{us}^\ell|^2\,f_{K^\pm}^2}{|\tilde V_{ud}^\ell|^2\,f_{\pi^\pm}^2}\left(1+\Delta^{K/\pi}_{\ell2} \right)
\label{eq:epslKpi}\\
=&\frac{|\tilde V_{us}^\ell|^2\,f_{K^\pm}^2}{|\tilde V_{ud}^\ell|^2\,f_{\pi^\pm}^2}\left(1-4(\eR^s-\eR^d)-
\frac{2B_0}{m_\ell}\left(\eP^{s\ell}-\eP^{d\ell}\right)\right)+\mathcal{O}(v^4/\Lambda^4)
\label{eq:deltarl1}
\end{align}
Notice that $R_\ell$ is not only sensitive to the NP-modified CKM elements $|\tilde V_{us}^\ell|/|\tilde V_{ud}^\ell|$ but also, 
and independently, to the right-handed or pseudoscalar operators. Last, we note that this result is in agreement with Ref.~\citep{Antonelli:2008jg} 
(Eq. 2.37).~\footnote{The relation between their WC and ours is the following: $c_{LL}^V=-1-\epsilon_L^*$, $c_{RL}^V=-\epsilon_R^*$, 
$c_{RR}^S=-(\epsilon_S^*+\eP^*)/2$, $c_{LR}^S=-(\epsilon_S^*-\eP^*)/2$ and $c_{RR}^T=-\epsilon_T^*$ (flavor indexes implicit). 
The remaining coefficients are zero in our EFT since they involve operators with right-handed neutrinos.} 

\subsection{$P_{\ell2\gamma}$}
\label{sec:Kl2gamma}

For the sake of completeness we discuss now briefly radiative pion and kaon decays, $P\to\ell\nu\gamma$ ($P_{\ell2\gamma}$), as NP probes. 
In addition to the QED correction (internal-bremsstrahlung) to the $P\to\ell\nu$ decay, we have the  
so-called ``structure-dependent'' terms which can be extracted separately from experiment~\cite{Bolotov:1990yq,Bychkov:2008ws,Ambrosino:2009aa}; 
in the SM they depend on $P\to\gamma$ hadronic form factors and some of them are not chirally suppressed~\cite{Bijnens:1992en,Cirigliano:2011ny}. 
The interest of $P_{\ell2\gamma}$ in the context of NP is that their kinematic distributions are sensitive to the tensor 
operator~\cite{Poblaguev:1990tv,Chizhov:1995wp}, which enter with new form factors~\cite{Mateu:2007tr} ($q=p-k$):
\begin{align}
\langle \gamma(k,\epsilon)|\bar u\sigma^{\mu\nu}\gamma_5D|P^-(p)\rangle=&\frac{e}{2}\,F^P_T(q^2)\,\left(\epsilon^\mu k^\nu-\epsilon^\nu k^\mu\right)\nonumber\\
&+G_T^P(q^2)\left[\epsilon\cdot p\left(p^\mu k^\nu-p^\nu k^\mu\right)+q\cdot p\left(\epsilon^\mu p^\nu-\epsilon^\nu p^\mu\right)\right], \label{eq:radff}
\end{align}
although $G_T^P$ is kinematically suppressed in the amplitude and can be neglected in first approximation. In fact, the PIBETA collaboration
has obtained a stringent constraint on these contributions in $\pi_{e2\gamma}$~\cite{Bychkov:2008ws} that, using the calculation of
$F^\pi_T$ obtained in Ref.~\cite{Mateu:2007tr}, leads to the bound~\cite{Bhattacharya:2011qm}:
\begin{align}
-1.2\times10^{-3}\leq\eT^{de}\leq1.36\times10^{-3}\hspace{0.5cm}\text{(90\% C.L.)}.\label{eq:eTboundrad} 
\end{align}
Similar experimental analyses have not been performed with $\pi_{\mu2\gamma}$ or $K_{\ell2\gamma}$ yet, where the experimental precision is not so high. 
Calculations of the kaon tensor form factors are also lacking, while the muonic channels are expected to be less sensitive
to tensor interactions, as they are dominated by the internal-bremsstrahlung part.

Finally, the structure-dependent terms also depend on the vector and axial $D\to u\ell\nu$ currents (e.g. the SM) and,
therefore, on $|\tilde V_{uD}^\ell|$ and $\eR^D$. However, in order to provide competitive values for these quantities one would
need high-accuracy data and LQCD results for the corresponding SM form factors.

\subsection{$K_{\ell3(\gamma)}$}
\label{sec:Kl3}

In the SM, the $K\to\pi\ell\nu$ decay amplitude depends on the hadronic matrix element~\cite{Antonelli:2008jg}:
\begin{eqnarray}
\langle\pi^{-}\left(  k\right)  |\bar{s}\gamma^{\mu}u|K^{0}\left(  p\right)
\rangle= P^\mu f_+(q^2) +q^\mu f_-(q^2),
\label{eq:ffsKl30}
\end{eqnarray}
where the $K^0\pi^-$ channel is taken as reference, $P=p+k$ and $q=p-k$. The $f_-(q^2)$ can be written in terms of $f_+(q^2)$ and the scalar form factor
$f_0(q^2)$ using the conservation of the vector current in QCD,
\begin{eqnarray}
&&\langle\pi^{-}\left(k\right) |\bar{s}u|K^{0}\left(p\right)
\rangle=-\frac{m_{K^0}^2-m_{\pi^\pm}^2}{m_s-m_u}f_0(q^2),\nonumber\\
&&f_0(q^2)=f_+(q^2)+\frac{q^2}{m_{K^0}^2-m_{\pi^\pm}^2}f_-(q^2),
\label{eq:f0} 
\end{eqnarray}
Finally, in presence of a tensor operator a new form factor appears~\cite{Antonelli:2008jg}:
\begin{align}
\langle\pi^{-}\left(k\right) |\bar{s}\sigma^{\mu\nu} u|K^{0}\left(p\right)\rangle=i\frac{p^\mu k^\nu-k^\mu p^\nu}{m_{K^0}}B_T(q^2).\label{eq:tensFF} 
\end{align}

\subsubsection{Kinematical distribution}

First let us briefly review the situation in the SM~\cite{Antonelli:2008jg,Antonelli:2010yf}. There are various methods proposed for the parametrization of 
the $q^2$ dependence of the form factors. The conventional one relies on a Taylor expansion,
\begin{eqnarray}
\bar{f}_{+,0}(q^2)=\frac{f_{+,0}(q^2)}{f_{+,0}(0)}=1+\lambda^\prime_{+,0}\frac{q^2}{m_\pi^2}+\frac{1}{2}\lambda^{\prime\prime}_{+,0}\left(\frac{q^2}{m_\pi^2}\right)^2+\ldots,\label{eq:FFsTaylor}
\end{eqnarray}
where the higher orders terms are negligible in the kinematic range of the decay, $q^2\in[m_\ell^2,\,(m_K-m_\pi)^2]$. These 
parameters are customarily fitted to the kinematic distributions of the decay (or Dalitz plot)~\cite{Yushchenko:2003xz,Yushchenko:2004zs,Lai:2004kb,Alexopoulos:2004sy,Ambrosino:2006gn,Ambrosino:2007ad,Lai:2007dx,Abouzaid:2009ry}, allowing for a calculation the phase-space integral (see next subsection).

The Taylor-expansion parametrization introduces a number of parameters which can not be always determined experimentally free of
ambiguities and more efficient parametrizations have been proposed~\cite{Jamin:2001zq,Bernard:2006gy,Bernard:2009zm,Antonelli:2008jg}, incorporating 
physical constraints to reduce the number of independent parameters. In particular, for $f_0(q^2)$ one can use a dispersive
representation \cite{Jamin:2001zq,Bernard:2006gy,Bernard:2009zm} that allows one to relate all its slope parameters to a single quantity that needs
to be measured and that is chosen to be $\log\,C$, where
\begin{equation}
C=\bar f_0(
m_K^2-m_\pi^2).
\end{equation}
Importantly, the value of this quantity can be determined very precisely in QCD using the Callan-Treiman theorem (CTT)~\cite{Callan:1966hu}
\begin{equation}
C_{\rm \scriptscriptstyle QCD}=\frac{f_K}{f_\pi}\frac{1}{f_+(0)}+\Delta_{\rm CT},\label{eq:CTth}
\end{equation}
where $\Delta_{\rm CT}=-0.0035(80)$ is a  small $\mathcal{O}(m_{u,d}/(4\pi f_\pi))$ correction calculated using ChPT~\cite{Gasser:1984ux,Bijnens:2007xa}.

It is also interesting to note that LQCD calculations of the $q^2$-dependence of the SM form factors have recently appeared~\cite{Carrasco:2016kpy}, 
although their precision is still smaller than the experimental determinations. Their inclusion in the future should be straightforward
and it should help obtaining a more precise $|\tilde V^\ell_{us}|$ determination and stronger NP bounds, while at the same time making the SM
calculations more robust.

Scalar and tensor operators modify the kinematic distribution and they should be determined together with the form factor parameters in
the fits to the Dalitz plots. First of all, their interference with the SM is proportional to the lepton mass
due to their chirality-flipping nature. A consequence of this is that the dependence on the corresponding WC for the 
electronic mode is, in very good approximation, quadratic and their kinematic distributions are SM-like at leading order of the EFT expansion.

In the muon channel, the effect of the scalar operator can be absorbed in the scalar form factor~\cite{Antonelli:2008jg}. This can be 
easily seen at the very amplitude level:
\begin{align}
\mathcal{M}(K^0\to\pi^-\mu\nu)&=-\frac{G_F\,\tilde V^{\mu*}_{us}}{\sqrt{2}}\left[
\left(\bar{u}\slashP(1-\gamma_5) v+
m_\mu\frac{m_{K^0}^2-m_{\pi^\pm}^2}{q^2}
\bar{u}(1+\gamma_5) v\right)f_+(q^2)\right.\nonumber\\
&\left.
-m_\mu\,\frac{m_{K^0}^2-m_{\pi^\pm}^2}{q^2}f_0(q^2)\left(1+\epsilon_S^{s\mu}\frac{q^2}{m_\mu(m_s-m_u)}\right)\bar{u}(1+\gamma_5) v+\ldots\right],\label{eq:ampKl3} 
\end{align}
where the dots correspond to the tensor contribution that we discuss below. Since this effect vanishes for $q^2=0$, it is easy to 
see that the whole effect of a scalar interaction ends up hidden in the $q^2$-dependence of the scalar form factor $f_0(q^2)$.
If precise values for $f_+(0)$ and $f_K/f_\pi$ are provided in QCD, the CTT gives a very accurate prediction of this form factor at $q^2=m_K^2-m_\pi^2$, which allows to separate $\epsilon_S^{s\mu}$ from the experimental measurement:
\begin{align}
\log C_{\rm exp}
&=\log \left[C_{\scriptscriptstyle{\rm QCD}} \left( 1+ \frac{m_K^2-m_\pi^2}{m_\mu(m_s-m_u)} \epsilon_S^{s\mu} \right)\right] \nonumber\\
&= \log C_{\scriptscriptstyle{\rm QCD}} + \frac{m_K^2-m_\pi^2}{m_\mu(m_s-m_u)} \epsilon_S^{s\mu}+\mathcal{O}(v^4/\Lambda^4)~.\label{eq:lnCtilde}
\end{align}

On the other hand, the tensor term can not be described by a simple re-definition of the SM contributions. This can be 
appreciated better by looking at the differential decay rate in terms $q^2$ and the angle $\theta$ defined in the $q$ rest frame by
the 3-momenta of the charged lepton and the one of the recoiling pion, $\vec k$:
\begin{align}
\frac{d\Gamma}{dq^2 d(\cos\theta)}&=\frac{G_F^2|\tilde V_{us}^\mu|^2}{128\pi^3}\,C_K\,S_{\rm ew}\,\left(1 + \delta^c+\delta^{c\mu}_{\rm
em}(q^2,\,\theta)\right)^2\frac{|\vec k|}{m_K^2}\left(1-\frac{m_\mu^2}{q^2}\right)^2\times\nonumber\\
&\left\{
\sin^2\theta\left|2m_K|\vec k|f_+(q^2)\left(1-\frac{2\epsilon_T^{s\mu}m_\mu}{m_K}\frac{ B_T(q^2)}{f_+(q^2)}\right)\right|^2\right.\nonumber\\
&\left.+m_\mu^2\left|\frac{2m_K |\vec k|}{\sqrt{q^2}} \cos\theta f_+(q^2)\left(1-\frac{2\epsilon_T^{s\mu}q^2}{m_\mu m_K}\frac{B_T(q^2)}{f_+(q^2)}\right)
+\frac{m_K^2-m_\pi^2}{\sqrt{q^2}}\tilde{f}_0(q^2)\right|^2\right\}
\,,\label{eq:Kl3diffrate}
\end{align}
where $C_K=1$ ($1/2$) for the neutral (charged) kaons, 
$\tilde{f}_0(q^2)=f_0(q^2)\left(1+\epsilon_S^{s\mu}\frac{q^2}{m_\mu(m_s-m_u)}\right)$ denotes the scalar form factor modified by NP, 
$\delta^{c\mu}_{\rm em}(q^2,\,\theta)$ are radiative corrections and $\delta^c$ is the isospin-breaking correction for the charged kaon channel, 
which can be obtained in ChPT~\cite{Cirigliano:2001mk,Ananthanarayan:2004qk,DescotesGenon:2005pw,Kastner:2008ch,Cirigliano:2008wn}. It is evident that the tensor operator 
introduces a characteristic dependence on $q^2$ and $\theta$
that is different from the SM. 

\subsubsection{Total rates}

The photon-inclusive $K_{\ell3}$ total decay rates can be written as~\cite{Antonelli:2008jg}
\begin{equation}
\Gamma(K_{\ell 3(\gamma)}) =
{G_F^2 m_K^5 \over 192 \pi^3}\, C_K\,
  S_{\rm ew}\,|\tilde{V}_{us}^{\ell}|^2 f_+(0)^2\,
I_K^\ell(\lambda_{+,0},\,\epsilon^{s\ell}_{S,T})\,\left(1 + \delta^c+\overline\delta^{c\ell}_{\rm
em}\right)^2\,\label{eq:Kl3rate},
\end{equation}
where $\overline\delta^{c\ell}_{\rm em}$ is the integrated radiative correction and $I_K^\ell(\lambda_{+,0},\epsilon^{s\ell}_{S,T})$ is the phase space integral, where $\lambda_{+,0}$ should be interpreted as a generic reference to the parameters describing the $q^2$ dependence of the form factors. Its expression is given by
\begin{align}
{ I}^\ell_K &= {I}_{K,0}^\ell - \epsilon^{s\ell}_T ~I_T^\ell +\mathcal{O}(v^4/\Lambda^4)~,
\label{eq:Kl3ratePSintegrals}\\
{ I}^\ell_{K,0} &= \frac{\textstyle 1}{\textstyle f_+(0)^2} 
\!\int \frac{dq^2}{m_K^2 } ~ \lambda^{3/2}(q^2) ~
\left( 1+ \frac{\textstyle m_{\ell}^2}{\textstyle 2q^2} \right) \left( 1 - \frac{\textstyle m_{\ell}^2} {q^2} \right)^2 \nonumber\\
&~~~\times \left(f_+(q^2)^2 + \frac{\textstyle 3m_{\ell}^2
\left(m_K^2-m_\pi^2\right)^2}{\textstyle \left( 2q^2+m_{\ell}^2 \right) m_K^4 \lambda(q^2)  } \  \tilde{f}_0(q^2)^2 \right), \nonumber
\nonumber\\
{ I}^\ell_T &= \frac{\textstyle 1}{\textstyle f_+(0)^2} \!\int  \frac{dq^2}{m_K^2 } ~ 
\lambda^{3/2}(q^2) ~ 
\frac{6~\textstyle m_\ell}{\textstyle m_K} 
\left( 1 - \frac{\textstyle m_{\ell}^2} {\textstyle q^2} \right)^2 B_T(q^2) {f}_+(q^2),
\nonumber
\end{align}
where $\lambda(q^2)=1 - 2 r_\pi + r_\pi^2 - 2 q^2/m_K^2 - 2 r_\pi q^2/m_K^2+ q^4/m_K^4$ and $r_\pi=m^2_\pi/m^2_K$. Let us notice that the tensor contribution to the total rate, ${ I}^\ell_T$, does not agree with the result shown in the 2008 Flavianet report~\cite{Antonelli:2008jg}.

To determine the total rates beyond the SM one first needs to perform a global fit of the form-factor parameters \textit{and} $\epsilon_T^{s\ell}$ 
(provided a value for $B_T$) to the kinematic distribution in Eq.~(\ref{eq:Kl3diffrate}). This requires a careful assessment of the
uncertainty introduced by the radiative corrections $\delta^{c\ell}_{\rm em}(q^2,\,\theta)$ which can introduce sizable corrections to the rates
in some regions of the phase space~\cite{Cirigliano:2008wn}. Besides the parameters corresponding to $f_+(q^2)$, for the muonic mode one should obtain from the fit 
(correlated) intervals for $\log\,C$ and $\epsilon_T^{s\mu}$. As it will be discussed in more detail below, for the electronic mode our framework must be pushed
beyond the leading order in the EFT expansion to search for $|\epsilon_S^{se}|^2$ and $|\epsilon_T^{se}|^2$.
With the final results of these fits one can now compute the phase-space integral $I_K^\ell$ in Eq.~(\ref{eq:Kl3rate}), which allows for a determination of 
$f_+(0)|\tilde V^\ell_{us}|$ from the total rate $\Gamma(K_{\ell3(\gamma)})$. 

Similarly to $P_{\ell 2(\gamma)}$, there is also a lepton-universality ratio in $K_{\ell3(\gamma)}$ constructed from the total rates in which most of the theoretical 
uncertainties cancel:
\begin{align}
r_{\mu e} &=\frac{\Gamma_{K_{\mu3}} I_{e3} \left(1+2\delta_{\rm em}^{Ke}\right) }{\Gamma_{K_{e3}} I_{\mu3} \left(1+2\delta_{\rm em}^{K\mu}\right)} 
={\frac{|\tilde V_{us}^\mu|^2}{|\tilde V_{us}^e|^2} =
1+2\Delta^s_L}+\mathcal{O}(v^4/\Lambda^4)
~,\label{eq:Kl3LUrate}
\end{align}
and that is only sensitive to the difference of left-handed $s\to u$ currents, $\Delta^s_L = \eL^{s\mu}-\eL^{se}$, up to $\mathcal{O}(v^2/\Lambda^2)$ due to Eq.~(\ref{eq:RHCuniversal}).

\subsection{Nuclear, neutron and hyperon $\beta$ decay}
\label{sec:betadec}

The semileptonic decays of nuclei, neutron and hyperons are mediated by the same effective 
Lagrangian as the (semi)leptonic pion and kaons decays. We summarize here the aspects of these 
decays that offer the strongest synergies. 

The most accurate value for $|\tilde V_{ud}^e|$ is obtained from superallowed \textit{nuclear} $\beta$ transitions, 
in an analysis that also sets the most stringent limits on the non-standard scalar Wilson coefficient $\epsilon_S^{de}$ (via the Fierz interference term 
$b_F$)~\cite{Hardy:2014qxa,Bhattacharya:2011qm}. 
Combining this $|\tilde V_{ud}^e|$ determination with the $|\tilde V_{us}^{e}|$ value obtained from $K_{e3(\gamma)}$ allows 
one to test CKM unitarity, which probes the following combination of WC~\cite{Cirigliano:2009wk}:
\begin{align}
|\tilde V_{ud}^e|^2+|\tilde V_{us}^e|^2&=1+\Delta_{\rm CKM}~,\nonumber\\
\Delta_{\rm CKM}=2|\tilde V_{ud}^e|^2(\eL^{de}+\eR^{d})+2|\tilde V_{us}^e|^2&(\eL^{se}+\eR^{s})-2\frac{\delta G_F}{G_F}+\mathcal{O}(v^4/\Lambda^4).\label{eq:CKMuni}
\end{align}
where we have neglected the contribution of $|\tilde{V}_{ub}^{e}|^2$ because its value is smaller than the current uncertainty in $\Delta_{\rm CKM}$~\cite{Agashe:2014kda}. 

At the hadron level, neutron and hyperon $\beta$ decays are weighed by different form factors. For the SM contributions we have~\cite{Weinberg:1958ut}:
\begin{align}
\bra{B_2 (p_2) } \bar{u} \gamma_\mu D \ket{B_1 (p_1)} 
&=
\bar{u}_2 (p_2)  \left[
f_1(q^2)  \,  \gamma_\mu    
+ \frac{f_2(q^2)}{M_1}   \, \sigma_{\mu \nu}   q^\nu  
+ \frac{f_3(q^2)}{M_1}   \,  q_\mu  
\right]  
 u_1 (p_1)~,  \nonumber\\
\bra{B_2 (p_2) } \bar{u} \gamma_\mu \gamma_5  D \ket{B_1 (p_1)} 
&=
\bar{u}_2 (p_2)  \left[
g_1(q^2)    \gamma_\mu    
+ \frac{g_{2} (q^2)}{M_1}   \sigma_{\mu \nu}   q^\nu  
+ 
\frac{g_{3} (q^2)}{M_1}   q_\mu  
\right]  \,\gamma_5  u_1 (p_1)~,
\label{eq:baryonFFs}
\end{align}
while the (pseudo)scalar and tensor operators introduce new form factors~\cite{Weinberg:1958ut,Bhattacharya:2011qm,Chang:2014iba}.
The normalization of the decays, $f_1(0)\,|\tilde V_{uD}^\ell|$, leads to independent $|\tilde V_{uD}^\ell|$ constraints once a theoretical value
for the ``vector charge'' of the transition, $f_1(0)$, is used (see e.g. Refs.~\cite{Cabibbo:2003ea,Mateu:2005wi,Sasaki:2012ne,Geng:2014efa,Shanahan:2015dka}). 

The only effects of the right-handed currents $\epsilon_R^D$ in these decays enter hidden in $|\tilde V_{uD}^\ell|$, \textit{cf.} Eq~(\ref{eq:VuDNPs}), 
and in the axial form factors, like the ``axial charge'' of the transition, $g_1\equiv g_1(0)$ (commonly denoted by 
$g_A$ in the case of the neutron decay)~\cite{Bhattacharya:2011qm,Chang:2014iba}:
\begin{eqnarray}
g_A^{\rm expt}=(1-2\eR^d)\,g_A~,\hspace{1cm}
g_1^{\rm expt}=(1-2\eR^s)\,g_1~.\label{eq:g1}
\end{eqnarray}
Thus, a bound on the right-handed current can be determined if any of the axial form factors is both measured and calculated in LQCD. 
This is indeed the case for $g_A$, which has been measured precisely~\cite{Mund:2012fq,Mendenhall:2012tz} and for which there are ongoing
LQCD efforts~\cite{Green:2014vxa}, with results in the physical point currently at the few-percent 
level~\cite{Horsley:2013ayv,Bali:2014nma,Abdel-Rehim:2015owa}.
As recently pointed out in Ref.~\cite{Chang:2014iba}, our knowledge for the hyperon decays is far less advanced both experimentally and theoretically. 

The nonstandard coefficients $\epsilon_{S,P,T}^{D\ell}$ modify not only the total rate but also the kinematic distributions 
and polarization observables of the $\beta$ decays~\cite{Cabibbo:2003cu,Bhattacharya:2011qm,Gonzalez-Alonso:2013ura,Chang:2014iba}. Strong bounds 
on $\epsilon_{S,T}^{de}$ have been obtained from global fits to various precise measurements in nuclear and neutron 
decays~\cite{Severijns:2006dr,Wauters:2013loa,Pattie:2013gka,Hardy:2014qxa,Gonzalez-Alonso:2013uqa}, whereas somewhat weaker (but still nontrivial) 
bounds are expected for the pseudo-scalar term $\eP^{de}$~\cite{Gonzalez-Alonso:2013ura}.
It is also worth noting that the muonic NP-modified CKM matrix element, $\tilde V_{ud}^\mu$, and WC, $\epsilon_{S,P,T}^{d\mu}$, 
cannot be determined from $\beta$ decays since the muon channels are kinematically forbidden. 

The analysis of these contributions can be extended to semileptonic hyperon decays~\cite{Chang:2014iba}. Similarly to $K_{\ell3(\gamma)}$, 
the chiral suppression of (pseudo)scalar and tensor operators implies that only the muonic case presents a non-negligible linear dependence on the 
WC. For instance, in Ref.~\cite{Chang:2014iba} the following lepton universality ratio was studied:
\begin{align}
R_{B_1B_2}=\frac{\Gamma(B_1\to B_2\mu\nu)}{\Gamma(B_1\to B_2 e\nu)}
=\frac{|\tilde V_{us}^{\mu}|^2}{|\tilde V_{us}^{e}|^2}(1+R_S\,\epsilon_S^{s\mu}+R_T\,\epsilon_T^{s\mu})
~,\label{eq:SHDNP}
\end{align}
where the coefficients $R_{S,T}$ depend of the decay channel~\cite{Chang:2014iba}. The NP contributions to $|\tilde V_{us}^{\mu}|/|\tilde V_{us}^{e}|$ are encoded in $\Delta_L^s$ which can be extracted independently from $K_{\ell 3}$ decays, \textit{cf.} Eq.~(\ref{eq:Kl3LUrate}), so that measuring $R_{B_1B_2}$ in different channels allows to set bounds on $\epsilon_{T,S}^{s\mu}$ at the few per-cent level, even though the old hyperon decay data set is used as input~\cite{Chang:2014iba}. New experiments and a comprehensive analysis of observables is needed to 
fully exploit the interesting degree of complementarity between hyperon and kaon decays. 

\section{Strategy for the global analyses}
\label{sec:strategy}

Having discussed all the ($CP$-averaged) observables appearing in $P_{\ell2(\gamma)}$ and $K_{\ell3(\gamma)}$, and its complementarity with baryon decays, 
we will now outline a strategy to take into account \textit{all} the information about NP one can extract from the experimental data, while summarizing also
the theoretical inputs needed.

Only three of the four $P_{\ell2(\gamma)}$ ratios discussed in Sec.~\ref{sec:Kl2} are independent and we need also to include in the analysis one total
rate (controlling the overall normalization of the rates). We choose $R_\pi$, $R_K$, $R_\mu$ and $\Gamma(K_{\mu2(\gamma)})$.
For the theoretical predictions we need $f_K^\pm/f_\pi^\pm$ 
and $f_{K^\pm}$ which are calculated accurately in LQCD, and the radiative corrections described in Sec.~\ref{sec:Kl2}. 
The output quantities obtained are: 
\begin{align}
\Big\{R_\pi,\,R_K,\,&R_{\mu},\,\Gamma(K_{\mu2(\gamma)})\Big\}\nonumber\\
&\Big\downarrow\hspace{0.5cm}{\scriptstyle R_\pi^{\rm th},\,R_K^{\rm th},f_{K^\pm}/f_{\pi^{\pm}},\,f_{K^{\pm}},\,\text{radiative corrections}}\nonumber\\
\Bigg\{\frac{|\tilde V^e_{ud}|^2}{|\tilde V^\mu_{ud}|^2}\left(1+\Delta^{\pi}_{e2/\mu2}\right),
\,\frac{|\tilde V^e_{us}|^2}{|\tilde V^\mu_{us}|^2}(1+&\Delta^K_{e2/\mu2}),\,
\frac{|\tilde V^\mu_{us}|^2}{|\tilde V^\mu_{ud}|^2}\left(1+\Delta^{K/\pi}_{\mu2}\right),\,
|\tilde V_{us}^\mu|^2\,\left(1+\Delta^K_{\mu2}\right)\Bigg\}\label{eq:schemaPl2}
\end{align}
where the $\Delta^X$ are the combinations of the WC in Eqs.~(\ref{eq:epsPl2ell},~\ref{eq:epsDemu},~\ref{eq:epslKpi}). 

For $K_{\ell3(\gamma)}$, we have three (CP-averaged) channels for electron and muon, namely $K^{\pm},K_L$ and $K_S$. Since
they are sensitive to the same short-distance physics, we can simply average over them (taking into account SM long-distance effects
that affect them differently). The comparison of the output obtained in different channels (e.g. $f_+(0)\,|\tilde V_{us}^\ell|$)
is a useful experimental crosscheck~\cite{Antonelli:2008jg,Antonelli:2010yf}, but it does not provide any NP constraint in the EFT framework.

First, the kinematic distributions have to be fitted to a parametrization of the form factors and, also to $\,\epsilon_T^{s\mu}$ using the LQCD
determination for $B_T(q^2)$.
For the muonic mode, one extracts $\epsilon_S^{s\mu}$ comparing the experimental determination of $\log\,C$ with 
the value given by the CTT theorem and the lattice calculations of $f_+(0)$ and $f_K/f_\pi$, \textit{viz.} Eqs.~(\ref{eq:CTth},~\ref{eq:lnCtilde}).

With the correlated results of these fits one calculates the spectral integrals $I_K^e$ and $I_K^\mu$, that are then used to extract 
$f_+(0)\,|\tilde V_{us}^e|$ and $r_{\mu e}$ from the electronic and muonic total rates,
\textit{viz.} eqs~(\ref{eq:Kl3rate},~\ref{eq:Kl3LUrate}), which then gives $|\tilde V_{us}^\ell|$ using as input the LQCD determination for $f_+(0)$. Schematically:
\begin{align}
\text{Kinematic}\,&\,\text{distributions}\nonumber\\
&\Big\downarrow\hspace{0.5cm}{\scriptstyle \text{radiative corrections}} \nonumber\\
\Big\{\lambda_{+}^{\prime},\lambda_{+}^{\prime\prime},\,\log\,C_{\rm exp}&,\,B_T(q^2)\,\epsilon_{T}^{s\mu}\Big\}\nonumber\\
&\Big\downarrow\hspace{0.5cm}{\scriptstyle \log C_{\scriptscriptstyle{\rm QCD}},~B_T(q^2),\,} \nonumber\\
\Big\{I_K^e,\,I_K^\mu,&\,\epsilon_{S}^{s\mu},\,\epsilon_{T}^{s\mu}\Big\}\nonumber\\
\Big\{r_{\mu e},\,\Gamma(K_{e3(\gamma)})\Big\}\xlongrightarrow{\hspace{2cm}}\hspace{0.3cm}&\Big\downarrow
\hspace{0.5cm}{\scriptstyle f_+(0),\,\text{radiative and isospin corrections}}\nonumber\\
\Big\{|\tilde V_{us}^e|,\,\Delta_L^s &,\,\epsilon_{S}^{s\mu},\,\epsilon_{T}^{s\mu}\Big\}\label{eq:schemaKl3}
\end{align}

Note that, in general, and except for $\epsilon_{S,T}^{s\mu}$, the global analysis of $P_{\ell2(\gamma)}$ and $K_{\ell3(\gamma)}$
does not allow to determine each WC separately, but only certain combinations of them. Baryon $\beta$ decays provide extra observables
that can help to disentangle most of them individually. For instance, including the determination of $|\tilde V_{ud}^e|$ allows one
to access $\Delta_{\rm{CKM}}$ via Eq.~(\ref{eq:CKMuni}) and the WC combination $\Delta^{K/\pi}_{\mu2} -2\Delta^d_L$ from $R_\mu$. This leads to a reinterpretation of the
classical $|V_{us}|-|V_{ud}|$ plot illustrating the consistency of $K_{\ell 3}$, $K_{\ell 2}/\pi_{\ell 2}$ and nuclear $\beta$ 
decays determinations of $|V_{ud}|$ and $|V_{us}|$~\cite{Marciano:2004uf}. In our EFT approach such test represents a powerful 
probe of the NP contribution $\Delta^{K/\pi}_{\mu2}-2\Delta^d_L$, whereas the additional consistency with the unitarity condition probes 
$\Delta_{\rm{CKM}}$. We will come back to this point in Sec.~\ref{sec:VudVusplot}.

Last but not least, the analysis of the nucleon and hyperon axial charges allows to extract 
$\epsilon_{R}^D$ ($D=d,s$), which, in turn, makes possible to set individual bounds on $\eP^{se}$, $\eP^{s\mu}$ 
and $\eP^{de}$ from the kaon fit output. Schematically:
\begin{align}
\Bigg\{
|\tilde V_{us}^e|,\,\Delta_L^s,\,\frac{|\tilde V^e_{ud}|^2}{|\tilde V^\mu_{ud}|^2}\left(1+\Delta^{\pi}_{e2/\mu2}\right),
\,\frac{|\tilde V^e_{us}|^2}{|\tilde V^\mu_{us}|^2}(1+&\Delta^K_{e2/\mu2}),\,
\frac{|\tilde V^\mu_{us}|^2}{|\tilde V^\mu_{ud}|^2}\left(1+\Delta^{K/\pi}_{\mu2}\right),\,
|\tilde V_{us}^\mu|^2\,\left(1+\Delta^K_{\mu2}\right)\Bigg\}\nonumber
\\
\Big\{|\tilde V_{ud}^e|,\, g^{\rm expt}_A,\,g^{\rm expt}_1 \Big\}\xlongrightarrow{\hspace{2cm}}\hspace{0.3cm}
&\Bigg\downarrow
\hspace{0.5cm}{\scriptstyle g_A^{\rm LQCD},\,g_1^{\rm LQCD},\,\text{CKM unitarity}}\nonumber
\\
\Bigg\{
|\tilde V_{ud}^e|,\,\Delta_{\rm CKM},\,\Delta_L^s,\,&\Delta^d_{LP},\,\eP^{se},\,\eP^{s\mu},\,\eP^{de},\,\epsilon_R^d,\,\epsilon_R^s\Bigg\}
\label{eq:schema3}
\end{align}

There are certain WC which cannot be determined individually using the low-energy data discussed thus far. For instance,
the lack of experimental input for a lepton-universality ratio $|\tilde V_{ud}^e|^2/|\tilde V_{ud}^\mu|^2$, precludes setting a bound on
the combination $\eL^{de}-\eL^{d\mu}$ separated from $\eP^{d\mu}$ in the combination:
\begin{align}
\Delta^d_{LP}= -\Delta^d_L+\frac{B_0}{m_\mu}\eP^{d\mu}\,=\, \eL^{de}-\eL^{d\mu}+\frac{B_0}{m_\mu}\eP^{d\mu},~\label{eq:WCsindet1}
\end{align}
obtained from $R_\pi$. Nonetheless, one gets access to these WC in muon-capture and inverse $\beta$-decay experiments, 
in which a few-percent experimental precision has been achieved. It would be interesting to investigate the potential
of these processes to provide independent bounds on NP in the context of the EFT approach described here.

In addition, the WC in the following combinations:
\begin{align}
\tilde{\Delta}_{\rm CKM}=& 2|\tilde V_{ud}^e|^2\eL^{de}+ 2|\tilde V_{us}^e|^2\eL^{se} - 2\frac{\delta G_F}{G_F} \approx  1.9\, \eL^{de} + 0.1\,\eL^{se} -2\frac{\delta G_F}{G_F} ,\nonumber\\
\Delta^s_L=&\epsilon_{L}^{s\mu}-\epsilon_{L}^{se}.\label{eq:WCsindet2}
\end{align}
are not determined individually. This can not be improved by adding other low-energy charged-current processes and the only way to access the orthogonal directions 
to these WC combinations is through the use of high-energy data, or neutral-current low-energy processes connected to those studied in this work due to 
the $SU(2)_L\times U(1)_Y$ symmetry in the EFT.

\begin{figure}[h!]
\centerline{\includegraphics[scale=1.01]{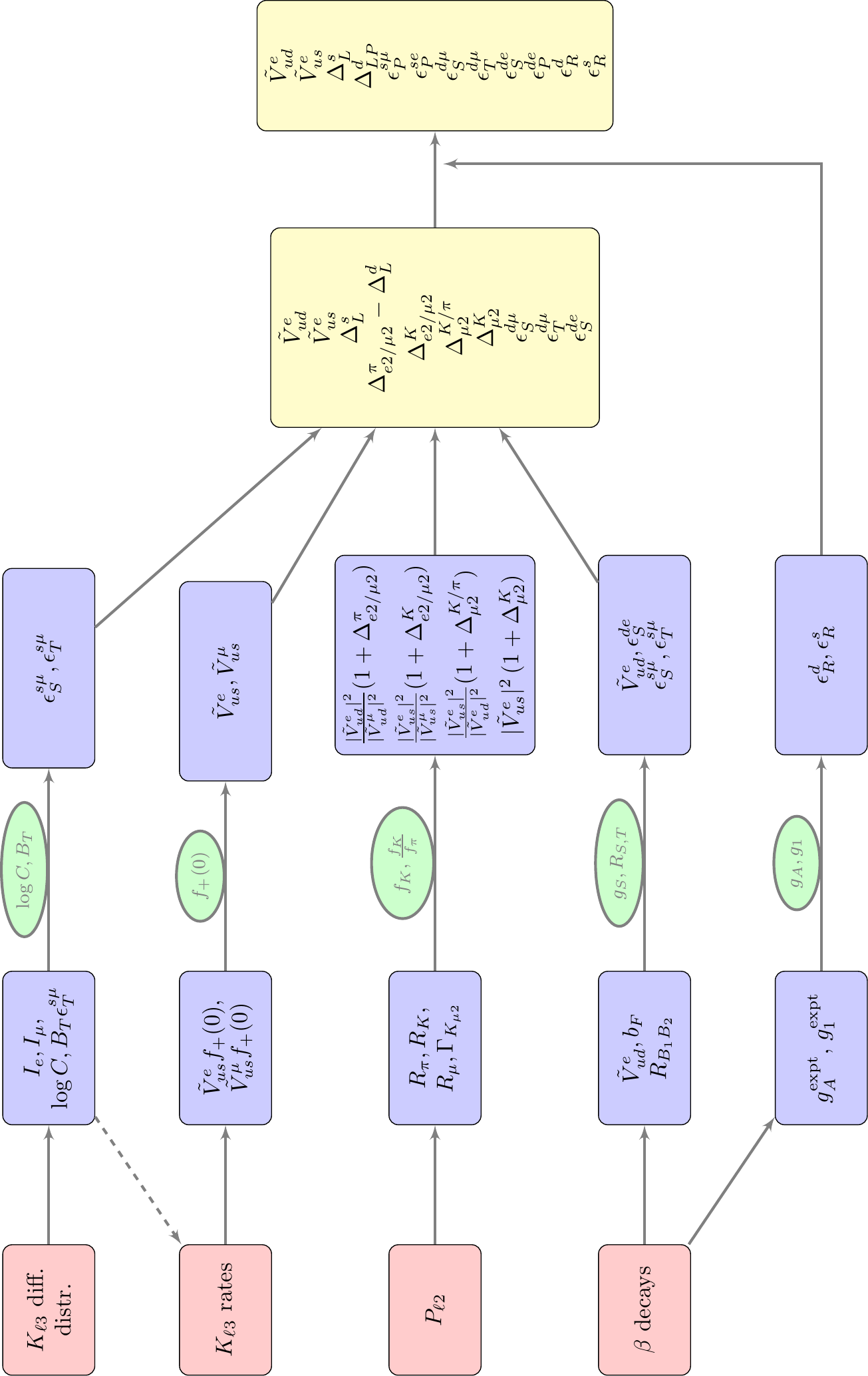}}
\caption{Flowchart.\label{fig:flowchart}}   
\end{figure}

In Fig.~\ref{fig:flowchart} we present a flowchart describing the correlation among different low-energy processes in a global (linearized) EFT analysis of NP in $D\to u\ell\nu$ transitions ($D=d,s$, $\ell=e,\mu$) and summarizing the different experimental and the theoretical inputs that are needed.

\subsection{Quadratic contributions of the WC}

In principle, it is possible to extend our analysis to include quadratic contributions of the WC to the observables,
although these count as $\mathcal{O}(v^4/\Lambda^4)$ in the EFT expansion and one needs to promote the global analysis 
to that order. In particular, the relation $\eR^{De}=\eR^{D\mu}$ would be violated by the interference of the SM with 
dimension-8 operators. For the sake of clarity, in this work we restrict ourselves to the few cases where the 
quadratic contributions can give the leading NP effects.

As discussed in Sec.~\ref{sec:Kl3}, this is the case of the WC $\epsilon_{S,T}^{se}$, which are not constrained 
in a linear fit to the (semi)leptonic kaon decay data, and whose quadratic terms represent the leading NP contributions to the $K_{e3}$ differential
distributions. Indeed, this has been used by the ISTRA~\cite{Yushchenko:2004zs} and NA48~\cite{Lai:2004kb} Collaborations 
to set bounds on those WC. However, none of these fits contain simultaneously the four relevant 
quantities, namely, the leading SM form-factors parameters $\lambda_+^\prime,\lambda_+^{\prime\prime}$ and both WC $\epsilon_{S,T}^{se}$, and the correlations are not given either.
The strongest bounds were obtained by ISTRA~\cite{Yushchenko:2004zs} in fits to $\lambda_+,\lambda_+^\prime$ and one of
the WC at a time, finding
\bea
\frac{f_S}{f_+(0)} &=& \frac{m_K^2-m_\pi^2}{2m_K(m_s-m_u)}\eS^{se} =- 0.004(^{+7}_{-6})(4)~,\\
\frac{f_T}{f_+(0)} &=& -2 \frac{B_T(0)}{f_+(0)}\eT^{se} = -0.012(21)(11)\label{eq:quadboundsISTRA}~,
\eea
where we have transformed the results in terms of the variables employed in that work, $f_T/f_+(0)$ and $f_T/f_+(0)$, into
those employed in this work.

On the other hand, as discussed in Sec.~\ref{sec:Kl2}, $\epsilon_P^{D\ell}$ contributes to $P_{\ell2}$ with a large helicity-enhancement that can make
their quadratic terms important too. In fact, this quadratic contribution allows for a second solution, different from zero, to the constraint that can
be obtained on the pseudoscalar WC. To be more explicit, this occurs when $1-(B_0/m_\ell) \eP^{D\ell}\simeq-1$, namely
for $\eP^{De}\sim4\times 10^{-4}$ or $\eP^{D\mu}\sim0.1$. In order to discard this other solution one would need an independent constraint 
on these WC that could be provided by $\beta$ decays~\cite{Gonzalez-Alonso:2013ura,Chang:2014iba}. Finally note that if we allow for a complex
CP-violating phase in the WC these two-fold degeneracies become circular in the complex plane of $\eP^{D\ell}$~\cite{Herczeg:1994ur,Bhattacharya:2011qm}.
\footnote{Any chiral-enhanced not-interfering term has the same consequences, e.g. operators with light right-handed~\cite{Herczeg:1994ur,Abada:2012mc,deGouvea:2015euy}
or ``wrong-flavor'' neutrinos~\cite{Bhattacharya:2011qm}.}

Finally, for completeness, we discuss the subleading operators ($D>6$) in the low-energy EFT, i.e. operators with derivatives neglected in the effective 
Lagrangian of Eq.~\eqref{eq:leff1}. These terms are corrections of order $(q/v)^n \lesssim (M_K/v)^n \sim (10^{-3})^n$ ($n\geq1$) 
with respect to the non-derivative terms. In the SM they are generated with $n=2$ at tree level (NLO terms in the W-propagator expansion) and 
are thus still unobservable in beta decays. An example of NP giving this type of contributions (with $n=1$) are the dipole-type ($D=6$) 
$SU(2)\times U(1)$-invariant operators such as $(\bar \ell \sigma^{\mu\nu} e) \tau^I \varphi W^I_{\mu\nu}$. Their effect will be then of
order $10^{-3}\times v^2/\Lambda^2$, which will be unobservable even for NP scales not much larger than the EW scale.

\section{Phenomenology}
\label{sec:Pheno}

\subsection{Inputs}
\subsubsection{Experimental}
\begin{table}[b]
\caption{Experimental data used in the analysis. See main text for more details about those values that do not have a reference in the table.
Additionally, the various masses and $\alpha_{\rm em}$ are taken from the PDG~\cite{Agashe:2014kda}.}
\begin{center}
\begin{tabular}{|c|c|c|}
\hline
$P_{\ell2(\gamma)}$&$K_{\ell3(\gamma)}$&Baryon $\beta$-decay
\\\hline
$R_\pi=1.2344(30)\times10^{-4}$~\cite{Aguilar-Arevalo:2015cdf}	
&~~$\left[\tilde V^e_{us}\,f_+(0)\right]_{\epsilon_T=0}=0.21649(44)$~~
&$|\tilde V_{ud}^e|=0.97451(38)$	
\\
$R_K =2.488(9)\times10^{-5}$~\cite{Agashe:2014kda}	
&~~$\left[\tilde V^\mu_{us}\,f_+(0)\right]_{\epsilon_T=0}=0.21667(54)$~~
&$b_f = -0.0028(26)$~\cite{Hardy:2014qxa}	
\\
$\Gamma(K_{\mu 2}) = 5.134(10)\times 10^7 s^{-1}$
&$\log C = 0.1985(70)$~\cite{Moulson:2014cra}	
&$g_A = −1.2723(23)$~\cite{Aguilar-Arevalo:2015cdf} 
\\
$\mbox{BR}(\pi_{\mu 2}) = 0.9998770(4)$~\cite{Agashe:2014kda}
&$-2\epsilon_T^{s\mu} \frac{B_T(0)}{f_+(0)} = -0.0007(71)$~\cite{Yushchenko:2003xz} & $\left.\frac{g_1}{f_1}\right|_{\Lambda p}=0.718(15)$~\cite{Agashe:2014kda}
\\
$\tau_{\pi^{\pm}} = 2.6033(5)  \times 10^{-8}~s $~\cite{Agashe:2014kda}
& &
\\
\hline
\end{tabular}
\end{center}
\label{tab:expinput}
\end{table}

We summarize in Table \ref{tab:expinput} the experimental values used for our analysis. Now we discuss some non-trivial aspects of them: 
\begin{itemize}
\item We calculate $\Gamma(K_{\mu 2})$ using the latest Flavianet results for the corresponding BR and the lifetime~\cite{Moulson:2014cra},
and taking into account their $10\%$ correlation~\cite{Moulson:private}.
\item {\bf $K_{\ell 3}$ shapes}: We take the value of $\log C$ from the latest Flavianet update~\cite{Moulson:2014cra} and the
tensor term for the muonic mode from the ISTRA+ analysis~\cite{Yushchenko:2003xz}. We assume these bounds will hold in a combined
fit, and we neglect the correlations of $\log C$, the tensor term and the phase-space integrals $I_{K,0}^\ell$. It should be straightforward to amend these limitations by the 
experimental collaborations.
\item {\bf $K_{\ell 3}$ rates}: We take the latest Flavianet results~\cite{Moulson:2014cra} for the product $|\tilde V^\ell_{us}|\,f_+(0)$ obtained from the channels $K_{Le3},K_{L\mu3},K_{Se3},K_{e3}^{\pm},K_{\mu3}^{\pm}$, and taking into account their correlations~\cite{Moulson:private} we average them to obtain the values $|\tilde V^e_{us}|\,f_+(0)$ and $|\tilde V^\mu_{us}|\,f_+(0)$ shown in Table \ref{tab:expinput}, which present a $+52\%$ correlation. 
Since the Flavianet extraction sets to zero the tensor term, these numbers are still missing the contribution from the tensor phase space 
integral, cf. Eqs.~\eqref{eq:Kl3rate}-\eqref{eq:Kl3ratePSintegrals}, which we denote in Table \ref{tab:expinput} with the subindex 
``$\epsilon_T=0$''. Combining these values with the bound on the tensor term, we find
\bea
\left(
\begin{array}{c}
 |\tilde V^e_{us}|\,f_+(0) \\
 |\tilde V^\mu_{us}|\,f_+(0)\\
 \eT^{s\mu} \frac{B_T(0)}{f_+(0)} \\
\end{array}
\right)
=
\left(
\begin{array}{c}
 0.21649(44) \\
 0.21670(59) \\
 -0.0007(71) \\
\end{array}
\right)~,~~
\rho=
\left(
\scriptsize{
\begin{array}{ccc}
 1. & 0.47 & 0. \\
 - & 1. & -0.42 \\
 - & - & 1. \\
\end{array}
}
\right)~,
\label{eq:Vusfp}
\eea
where we have simply used that the total phase space integral is given by
\bea
I_K^\mu
\approx I_{K,0}^\mu \left( 1 - 0.40 \,\eT^{s\mu} \right)~,
\eea
using the Flavianet determination of $I_{K,0}^\mu$ and $B_T(0)/f_+(0)$ from Ref.~\cite{Baum:2011rm}.

\item {\bf Nuclear $\beta$ decays:} Current studies of superallowed nuclear transitions contain a SM analysis where $V_{ud}$ is extracted, 
finding $|\tilde V_{ud}^e|=0.97417(21)$, and a NP analysis where the Fierz term $b_F$ is bounded, $b_F = -0.0028(26)$~\cite{Hardy:2014qxa}. 
In our framework we need a combined extraction of both quantities. Assuming a Gaussian $\chi^2$ we can reconstruct the outcome of such a 
fit. For this reconstruction we use that the minimum of the 2-parameter fit is $\{{\cal F}t,b_F\} = \{3070.1\,s,-0.0028\}$~\cite{HT:private}, 
where ${\cal F}t$ is the so-called ``corrected'' ${\cal F}t$ value, where different nuclear-structure and transition-dependent radiative corrections 
have been subtracted (see e.g. Ref.~\cite{Hardy:2014qxa}). We obtain:
\bea
\left(
\begin{array}{c}
 |\tilde V^e_{ud}| \\
 b_F \\
\end{array}
\right)
=
\left(
\begin{array}{c}
 0.97451(38) \\
 -0.0028(26) \\
\end{array}
\right)~,~~
\rho=
\left(
\scriptsize{
\begin{array}{cc}
 1. & -0.83 \\
 - & 1. \\
\end{array}
}
\right)~.
\label{eq:Vude}
\eea
These results should be taken with caution, keeping in mind their reconstructed nature. Ideally, in the future they will be given in this format, 
where it is trivial to recover the SM limit setting $b_F=0$. Let us remind that $b_F=-2\,g_S\,\eS^{de}$, where $g_S$ is the corresponding scalar form factor (see e.g. Ref.~\cite{Gonzalez-Alonso:2013ura}).
\item
{\bf Neutron $\beta$ decays:} We use the PDG average for the axial charge $g_A$. Let us notice that this determination, which comes typically from the measurement of
the neutron $\beta$ asymmetry $A$, assumes the SM is correct, whereas in our EFT framework it provides a $g_A$ value modified by scalar
and tensor interactions (in addition to $\epsilon_R^d$). However, given the current bounds on $\epsilon_{S,T}^{de}$, these effects can
be neglected in comparison with the error of the lattice $g_A$ determination, which by far limits the $\epsilon_R^d$ bound. And the same
applies to the hyperon axial charge $g_1$.
\item {\bf Hyperon $\beta$ decays:} In contrast to $g_A$, the axial-charges in the hyperons decays, $g_1$, are measured with a relative
uncertainty not better than $2\%$~\cite{Agashe:2014kda}. The best precision is achieved in $\Lambda\to p e^-\bar \nu$ that we will use as a reference for the extraction of $\epsilon_R^s$ in the fits. As discussed in Sec.~\ref{sec:betadec}, the hyperon decays also provide independent limits on
the scalar and tensor muonic WC from the lepton-universality ratio in Eq.~(\ref{eq:SHDNP}). The current bound on the tensor 
WC, $\eT^{s\mu}=-0.017(20)$ ($1\sigma$)~\cite{Chang:2014iba}\footnote{Here we take into account that the $\eT^{s\mu}$ definition of 
Ref.~\cite{Chang:2014iba} has a minus sign difference with the definition used in this work.}, is only 4 times less precise than the one obtained from the shapes of
$K_{\mu3}$ and does not depend on the assumptions described above for this mode. 
\end{itemize}

\subsubsection{Theoretical}
\begin{table}[b]
\caption{Theory inputs. See main text for more details about those values that do not have a reference in the table. The scale/scheme-dependent quantities are given in the $\overline{MS}$ 
at $\mu=2$ GeV.
}
\begin{center}
\begin{tabular}{|c|c|c|}
\hline
$P_{\ell2(\gamma)}$&$K_{\ell3(\gamma)}$&Baryon $\beta$-decay
\\\hline
$R_\pi^{\rm SM}=1.2352(1)\times10^{-4}$~\cite{Cirigliano:2007ga}	
&$f_+(0)=0.9661(32)$~\cite{Aoki:2013ldr}	
&$g_S=1.02(11)$~\cite{Gonzalez-Alonso:2013ura}
\\
$R_K^{\rm SM}=2.477(1)\times10^{-5}$~~\cite{Cirigliano:2007ga}	
&$\Delta_{\rm CT} = -0.0035(80)$~\cite{Gasser:1984ux,Antonelli:2010yf}	
&$g_A=1.24(4)$~\cite{Horsley:2013ayv}	
\\
$f_{K^\pm}/f_{\pi^\pm}=1.192(5)$~\cite{Aoki:2013ldr}
&$\frac{B_T(0)}{f_+(0)} = 0.68(3)$~\cite{Baum:2011rm}
&$\left.\frac{g_1}{f_1}\right|_{\Lambda p}=0.72(7)$
\\
$f_{K^\pm}=154.3(0.4)(2.8)$  MeV~\cite{Bazavov:2009bb} &	&
\\
$\delta_{\rm em}^{K\mu}-\delta_{\rm em}^{\pi\mu}=-0.0069(17)$ && \\
$\delta_{\rm em}^{K\mu}=0.0121(32)$ && \\
\hline
\end{tabular}
\end{center}
\label{tab:thinput}
\end{table}

We summarize in Table \ref{tab:thinput} the theory input for our analysis. Some comments are in order: 
\begin{itemize}
\item Using the expression in Eq.~(\ref{eq:Kl2rad}), we find $\delta_{\rm em}^{K\mu}=0.0121(12)_{c_1^K}(30)_{e^2p^4}$ for
the EM corrections to $K_{\mu 2(\gamma)}$ at order $e^2p^2$ in the chiral expansion. The first error comes from the 
uncertainty in the hadronic structure constant $c_1^K=-1.98(50)$~\cite{DescotesGenon:2005pw,Cirigliano:2007ga} at 
order $e^2p^2$ and the second error (25\% of the central value)
is an estimate of corrections due to higher order terms. Likewise we find $\delta_{\rm em}^{K\mu}-\delta_{\rm em}^{\pi\mu}=-0.0069(17)_{e^2p^4}$,
which is free of $c_1^P$ uncertainties at this order.
\item We use FLAG averages (with $N_f=2+1$)~\cite{Aoki:2013ldr} for both $f_{K^\pm}/f_{\pi^\pm}$~\cite{Follana:2007uv,Durr:2010hr,Arthur:2012yc} and $f_+(0)$~\cite{Bazavov:2012cd,Boyle:2013gsa}.
\item The experimental value of $f_\pi$ is often used to set the scale in the LQCD calculations. 
Within the SMEFT setup this is not convenient because one propagates the NP contribution $\Delta^{\pi}_{\mu2}$ onto all the 
dimensionful quantities determined thereafter in the lattice. Namely, the corresponding determination of $f_K$ makes $\Gamma(K_{\mu2(\gamma)})$ not sensitive to $\Delta^K_{\mu2}$, but only  to 
$\Delta^{K/\pi}_{\mu2}$, i.e. the direction already probed by the $R_{\mu}$ ratio. Thus, it is better to use determinations 
where an observable dominated by strong dynamics is used to set the scale. Among the determinations passing the FLAG
requirements~\cite{Follana:2007uv,Bazavov:2009bb,Arthur:2012yc,Aoki:2013ldr} we choose for our fit the MILC09 calculation as it already includes
the isospin corrections, $f_{K^\pm}=154.3(0.4)(2.8)$ MeV~\cite{Bazavov:2009bb}. 
Notice that this caveat also holds in a global SM analysis of $K_{\ell2}$ and $\pi_{\ell2}$ data, since using the experimental value of $f_\pi$ 
to set the QCD scale entails the loss of one of the experimental inputs.
\item Using the values in the Table \ref{tab:thinput} we find $\log C_{\rm \scriptscriptstyle QCD} = 0.2073(84)$. The correlation between this number
and the quantities $f_{K^\pm}/f_{\pi^\pm}$ and $f_+(0)$ is taken into account.
\item There are a few recent $N_f=2$ LQCD calculations of axial charge of the nucleon at the physical point~\cite{Horsley:2013ayv,Bali:2014nma,Abdel-Rehim:2015owa}. We use $g_A=1.24(4)$~\cite{Horsley:2013ayv}.
\item There are no computations of the axial charges for the semileptonic $\Delta S=1$ hyperon decays in the lattice yet, although pioneering calculations of the 
$\Delta S=0$ ones of the $\Sigma$ and $\Xi$ baryons have been reported~\cite{Lin:2007ap,Gockeler:2011ze,Cooke:2013qqa}. 
Taking the results in Ref.~\cite{Gockeler:2011ze} (and $g_A$ from Ref.~\cite{Horsley:2013ayv}), we obtain $g_\Sigma=0.91(4)$ and $g_\Xi=-0.25(3)$. These can be connected 
to the $\Delta S=1$ couplings using $SU(3)_F$, which is known to work empirically at few-percent accuracy for these quantities 
(see discussions in Refs.~\cite{Cabibbo:2003cu,Ledwig:2014rfa}). We obtain $\left.g_1/f_1\right|_{\Lambda p}=0.72(7)$, where 
we have conservatively estimated $SU(3)$-breaking corrections by a 10\%. Needless to say that the situation could be improved 
with direct LQCD calculations of these couplings.
\end{itemize}

\subsection{Fit}

Using the experimental and theoretical inputs listed in Tables~\ref{tab:expinput} and~\ref{tab:thinput}, and treating all errors as Gaussian, we perform a
standard $\chi^2$ fit, keeping only linear terms in $v^2/\Lambda^2$ in the theoretical expressions. The results are:
\bea
\left(
\begin{array}{c}
 \tilde{V}_{ud}^{e} \\
 \tilde{V}_{us}^{e} \\
 \Delta_L^s \\
 \Delta^d_{LP} \\
 \epsilon_P^{de} \\
 \epsilon_R^d\\
 \epsilon_P^{se} \\
 \epsilon_P^{s\mu} \\
 \epsilon_R^s \\
 \epsilon_S^{s\mu} \\
 \epsilon_T^{s\mu} \\
 \epsilon_S^{de} \\
\end{array}
\right)
=
\left(
\begin{array}{c}
 0.97451\pm 0.00038 \\
 0.22408\pm 0.00087 \\
 1.0\pm 2.5 \\
 1.9\pm 3.8 \\
 4.0\pm 7.8 \\
 -1.3\pm 1.7 \\
 -0.4\pm 2.1 \\
 -0.7\pm 4.3 \\
 0.1\pm 5.0 \\
 -3.9\pm 4.9 \\
 0.5\pm 5.2 \\
 1.4\pm 1.3 \\
\end{array}
\right)\times 10^{\wedge}\left(
\begin{array}{c}
 0 \\
 0 \\
 -3 \\
 -2 \\
 -6 \\
 -2 \\
 -5 \\
 -3 \\
 -2 \\
 -4 \\
 -3 \\
 -3 \\
\end{array}
\right)~,
\eea
in the $\overline{MS}$ scheme at $\mu=2$ GeV. 
Let us remind the reader that $\Delta_L^s = \epsilon_{L}^{s\mu}-\epsilon_{L}^{se} $ and 
$\Delta^d_{LP}= \eL^{de}-\eL^{d\mu}+\frac{B_0}{m_\mu}\eP^{d\mu} \,\approx \eL^{de}-\eL^{d\mu}+24\eP^{d\mu}$.

The correlation matrix is given by:
\bea
\rho=
\left(
\scriptsize{
\begin{array}{cccccccccccc}
 1. & 0. & 0. & 0.01 & 0.01 & 0. & 0. & 0. & 0. & 0. & 0. & 0.82 \\
 - & 1. & -0.16 & 0. & 0. & 0. & 0.04 & 0.04 & 0. & -0.26 & 0. & 0. \\
 - & - & 1. & 0. & 0. & 0. & -0.01 & 0.02 & 0. & 0. & 0.46 & 0. \\
 - & - & - & 1. & 0.9995 & -0.87 & 0.09 & 0.09 & 0. & 0.04 & 0. & 0.01 \\
 - & - & - & - & 1. & -0.87 & 0.09 & 0.09 & 0. & 0.04 & 0. & 0.01 \\
 - & - & - & - & - & 1. & 0. & 0. & 0. & 0. & 0. & 0. \\
 - & - & - & - & - & - & 1. & 0.9995 & -0.98 & -0.01 & 0. & 0. \\
 - & - & - & - & - & - & - & 1. & -0.98 & -0.01 & 0.01 & 0. \\
 - & - & - & - & - & - & - & - & 1. & 0. & 0. & 0. \\
 - & - & - & - & - & - & - & - & - & 1. & 0. & 0. \\
 - & - & - & - & - & - & - & - & - & - & 1. & 0. \\
 - & - & - & - & - & - & - & - & - & - & - & 1. \\
 \end{array}
}
\right)~.
\eea

Exploiting now the unitarity of the CKM matrix as explained in Sec.~\ref{sec:strategy}, trading  $|\tilde{V}_{us}^{e}|$ for 
$\Delta_{\rm CKM}$, we obtain:
\bea
 \Delta_{\rm CKM} 
 &=& 2|V_{ud}^e|^2(\eL^{de}+\eR^{d})+2|V_{us}|^2(\eL^{se}+\eR^{s})-2\frac{\delta G_F}{G_F}\nonumber\\
 &=&-(1.2\pm 8.4)\times 10^{-4}~,\nonumber\\
 \rho_{2i} &=&
 \left(
\scriptsize{
\begin{array}{cccccccccccc}
 0.88	&	1. & -0.07 & 0.01 & 0.01 & 0. & 0.02 & 0.02 & 0. & -0.12 & 0. & 0.73 \\
\end{array}
}
\right)~.
\eea
%

We observe good agreement with the SM, with marginalized limits varying from the $10^{-5}$ level for the pseudoscalar couplings
in the electronic channel (due to the chiral enhancement) to the per-cent level for the right-handed couplings 
(due to the limited lattice precision in the axial-vector form factors).

We observe also that the combinations of WC $\{\Delta^d_{LP},\,\epsilon_P^{de}\}$ and $\{\epsilon_P^{se},\,\epsilon_P^{s\mu} \}$
are highly correlated, which simply reflects the fact that the specific combination of them that appears in $R_\pi$ and $R_K$ respectively
is much more constrained than the individual WC. This is illustrated by the limits obtained when the only non-zero NP couplings
are the pseudoscalar couplings in the electronic channel:
\bea
 \epsilon_P^{de} = (0.7\pm 2.5)\times 10^{-7}~,\\
 \epsilon_P^{se} = -(4.5\pm 3.7)\times 10^{-7}~.
\eea
Such strong bounds can be the result of a very high NP scale, $\Lambda \sim v / \sqrt{\epsilon} \sim {\cal O}(500)$ TeV, or a non-trivial structure in lepton-flavor space, such as $\eP^{D\ell}\sim m_\ell\,\eP^{D}$. The latter case naturally
follows in models with extra Higgs doublets~\cite{Antonelli:2008jg} or, model-independently, from scalar four-fermion operators with 
Minimal Flavor Violation (MFV)~\cite{Chivukula:1987py,Hall:1990ac,Buras:2000dm,D'Ambrosio:2002ex,Cirigliano:2005ck} as in Ref.~\cite{Alonso:2015sja}. 

We complete the numerical discussion with the uncorrelated bounds that are obtained from $\pi_{e2\gamma}$, Eq.~(\ref{eq:eTboundrad}), and including quadratic effects 
in the Dalitz plots of $K_{e3}$, Eq.~(\ref{eq:quadboundsISTRA}): 
\begin{align}
&\eT^{de}=(0.1\pm0.8)\times10^{-3},\nonumber\\
&\eS^{se}=(-1.6\pm3.3)\times10^{-3},\nonumber\\
&\eT^{se}=(0.9\pm1.8)\times10^{-2},\label{eq:ancillbounds}
\end{align}
in the $\overline{MS}$ scheme at $\mu=2$ GeV. Here we have used the result in Table~\ref{tab:thinput} for $B_T(0)/f_+(0)$.

\subsubsection{The $V_{ud}-V_{us}$ plane revamped}
\label{sec:VudVusplot}

An application that particularly highlights the virtues of the EFT framework developed in this work 
is the $|V_{us}|-|V_{ud}|$ plot that illustrates the consistency of $K_{\ell 3}$, $K_{\ell 2}/\pi_{\ell 2}$ and
nuclear-$\beta$-decay determinations~\cite{Marciano:2004uf}. Interestingly enough, although the 
values of $|V_{ud}|$ and $|V_{us}|$ currently extracted from nuclear $\beta$ decays an $K_{\ell3}$ 
are in perfect agreement with unitarity, a small ``misalignment'' with the $K_{\ell2}/\pi_{\ell2}$ bound is observed~\cite{Aoki:2013ldr,Moulson:2014cra,Rosner:2015wva}.

In the general EFT setup, and if we focus on the electronic channel, this plot represents a projection of the global fit discussed earlier into
the $|\tilde V^e_{ud}|-|\tilde V^e_{us}|$ plane. The NP can manifest either as a violation of CKM unitarity or, precisely, as this misalignment of the
bound on $|\tilde V^e_{us}|/|\tilde V^e_{ud}|$ from $K_{e2}/\pi_{e2}$ with respect to the intersection of the other two bounds from $\beta$ decays
($|\tilde V^e_{ud}|$) and $K_{e3}$ ($|\tilde V^e_{us}|$). The former case corresponds to the bound on $\Delta_{\rm CKM}$ obtained above, whereas
the latter probes the combination of Wilson Coefficients $\Delta^{K/\pi}_e/2 = -2(\eR^s-\eR^d)- \frac{B_0}{m_e}\left(\eP^{se}-\eP^{de}\right)$, 
\textit{c.f.} Eqs.~(\ref{eq:epslKpi}-\ref{eq:deltarl1}). Hence, the right-handed and pseudoscalar contributions change the slope of the diagonal constraint obtained from $K_{e2}/\pi_{e2}$.

\begin{figure}[h!]
\centerline{\includegraphics[scale=0.40]{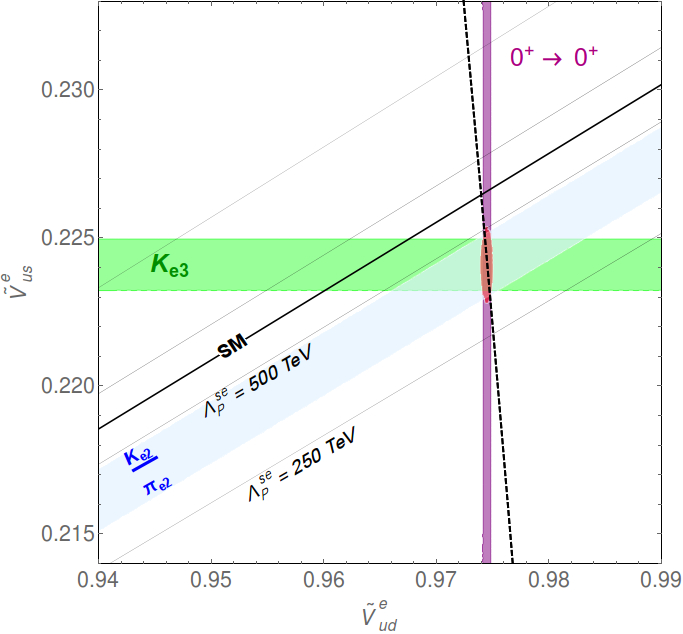}}
\caption{1$\sigma$ regions for $|\tilde V_{ud}^e|$ and $|\tilde V_{us}^e|$ from $K_{e3}$ (horizontal band) and nuclear $\beta$ decays 
(vertical band). We also plot the 1$\sigma$ region given by the ratio $\Gamma(K^\pm_{e2(\gamma)})/\Gamma(\pi^\pm_{e2(\gamma)})$ (diagonal band)
assuming a NP contribution $\Delta^{K/\pi}_{e2}\simeq0.02$, along with the lines corresponding to different NP effective scales $\Lambda_P^{se} \approx (V_{us} \epsilon_P^{se})^{-1/2}v$. 
The dashed black line shows the CKM unitarity constraint.
\label{fig:VudVus}}   
\end{figure}

In Fig.~\ref{fig:VudVus} we show current experimental constraints on the 
$|\tilde V^e_{ud}|-|\tilde V^e_{us}|$ plane, where we have added a NP contribution $\Delta^{K/\pi}_{e2}\simeq0.02$ needed to perfectly align the band
from $K_{e2}/\pi_{e2}$ with those from $\beta$-decays and $K_{e3}$. For illustration, we also show the diagonal lines corresponding 
to a NP contribution in $\Delta^{K/\pi}_{e2}$ if it was of the type $\epsilon_P^{se}$ for different values of the corresponding effective scale $\Lambda^{se}_P$.  

Nonetheless, this effect has a small significance, as reflected by the consistency of the data with the SM in our global fit discussed in the 
previous section. We see that this precise test of the SM, obtained thanks to the small experimental and theoretical uncertainties achieved 
in these processes, currently allows one to probe ${\cal O}(100)$ TeV scales.

Let us stress that Fig.~\ref{fig:VudVus} is obtained from our global fit, 
with all NP terms present, which makes the horizontal and vertical error bands wider. The traditional $V_{ud}-V_{us}$ plot is 
recovered if the only NP terms present are those probed in this plot, i.e. $\Delta_{\rm CKM}$ and $\Delta^{K/\pi}_{\mu2}$. That allows 
one to combine $K_{e3}$ and $K_{\mu 3}$ SM extractions of $\tilde{V}_{us}$ and to use the SM analysis of Ref.~\cite{Hardy:2014qxa} for $V_{ud}$, 
\text{c.f.} Eq.~\eqref{eq:Vude}. In that limit, the sensitivity to $\Delta_{\rm CKM}$ and $\Delta^{K/\pi}_{\mu2}$ is of course stronger, and a larger 
(though still not significant) tension arises in the plot~\cite{Aoki:2013ldr,Moulson:2014cra,Rosner:2015wva}.

Needless to say, one could have plotted instead the bound obtained from $K_{\mu2}/\pi_{\mu2}$, with the only difference that the
combination of WC probed in that case is longer, involving also left-handed $\epsilon_i$ to connect $\tilde V_{uD}^\mu$ with $\tilde V_{uD}^e$.
And the same applies e.g. for the $K_{\mu2}$ extraction.

\subsection{Minimal Flavor Violation and SM limits}

\begin{figure}[h!]
\centerline{\includegraphics[scale=0.45]{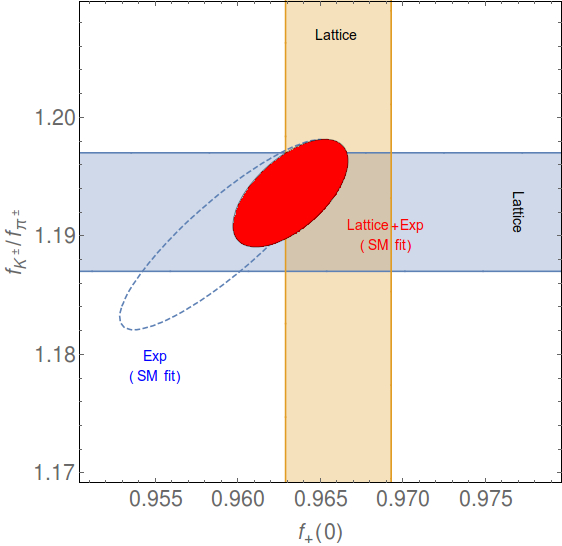}}
\caption{68\% C.L. regions for $f_{K^\pm}/f_{\pi^\pm}$ and $f_+(0)$ using only LQCD~\cite{Aoki:2013ldr}, only experimental data (in a SM fit) or using both.
\label{fig:FFplot}}   
\end{figure}

If the flavor symmetry $U(3)^5$ is respected, all NP terms vanish except those contaminating the CKM matrix elements, which in this
case become lepton-independent~\cite{Cirigliano:2009wk}. This NP flavor structure occurs if flavor breaking is suppressed by a mechanism such as 
Minimal Flavor Violation. 
Thus, the MFV analysis and the SM one (without imposing CKM unitarity) are equivalent. In this case we find:
\bea
\left(
\begin{array}{c}
 \tilde{V}_{ud} \\
 \tilde{V}_{us}\\
\end{array}
\right)
=
\left(
\begin{array}{c}
 0.97416(21) \\
 0.22484(64) \\
\end{array}
\right)~,~~
\rho=
\left(
\begin{array}{cc}
 1. & 0.03 \\
 - & 1. \\
\end{array}
\right)~,
\eea
where the tildes in the left-hand side apply only to the MFV case. The only NP probe left is then the CKM unitarity test~\cite{Cirigliano:2009wk}:
\bea
\Delta_{\rm CKM} = -(4.6\pm 5.2)\times 10^{-4}~.
\label{eq:deltackm}
\eea
In a SM analysis where the CKM unitarity is imposed, this NP term is set to zero, reducing the error in the matrix elements:
\bea
|V_{ud}| = 0.97432(12)~~~~\mbox{or equivalently}~~~~|V_{us}| = 0.2252(5).
\eea
Last but not least, we stress that our fit contains also the various QCD quantities as outputs. In the general EFT case,
they are trivially equal to their lattice QCD values that we use as inputs, since the fit is not overdetermined. On the other
hand, this is not the case in the SM limit, where experimental data do complement the lattice calculations~\cite{Antonelli:2010yf}, giving
\bea
\left(
\begin{array}{c}
 f_{K^\pm}\\
 f_{K^\pm}/f_{\pi^\pm}\\
 f_+(0) \\
\end{array}
\right)
=
\left(
\begin{array}{c}
 155.62(44) \mbox{MeV}\\
 1.1936(30) \\
 0.9632(23) \\
\end{array}
\right)~,~~
\rho=
\left(
\begin{array}{ccc}
 1. & 0.78 &0.56\\
 - & 1. &0.64\\
 - & - & 1.\\
\end{array}
\right)~.
\label{eq:FFfit}
\eea
Fig.~\ref{fig:FFplot} shows these results, and compare them with the bounds obtained using only LQCD~\cite{Aoki:2013ldr} or only experimental data, finding a good agreement.

The framework developed in this work is explicitly designed to make use of all the experimental information available in (semi)leptonic 
kaon and pion decays. In addition to the usual input of any analysis where $V_{ud}$ and $V_{us}$ are extracted (that is,  $K_{\ell3}$, 
$K_{\mu2}/\pi_{\mu2}$ and $\beta$ decays), our fit takes into account two more pieces of information. First, it includes the $K_{\mu2(\gamma)}$
rate that, along with a lattice determination for $f_{K^\pm}$, offers an additional handle on $V_{us}$. Secondly, we incorporate the CTT constraint 
in our fit, which, combined with the experimental determination of $\log C$, provides an extra handle on $f_{K^\pm}/f_{\pi^\pm}$ and $f_+(0)$;
namely $f_{K^\pm}/f_{\pi^\pm} \times 1/f_+(0) = 1.223(12)$. Indeed, this is part of the information that made possible the determinations shown
in Eq.~\eqref{eq:FFfit}. Although the CTT has been used before as an 
LQCD/ChPT check and as a NP probe 
(see e.g. Ref~\cite{Antonelli:2010yf}), it was not taken into account in the $V_{uD}$ extractions.

\begin{table}[h]
\caption{Comparison with other SM analyses (without imposing CKM unitarity and using $N_f=2+1$ lattice calculations).}.
\begin{center}
\begin{tabular}{ccccc}
\hline
Analysis ~~&~~ $V_{us}$ ~~~~~~& ~~Data ~~&~~ Form Factors~~ &~~ $K_{\mu2(\gamma)}$ and CTT
\\\hline\hline
This work &	 0.22484(64)	&	2014~\cite{Moulson:2014cra}	&	2013~\cite{Aoki:2013ldr} &	yes
\\\hline
Moulson'2014~\cite{Moulson:2014cra} &	 0.2248(7) &	2014~\cite{Moulson:2014cra}	&	2013~\cite{Aoki:2013ldr}	&	no
\\
(our code) &	 0.2248(7) &&
\\\hline
FLAG'2013~\cite{Aoki:2013ldr} &	 0.2247(7) &	2010~\cite{Antonelli:2010yf}	&	2013~\cite{Aoki:2013ldr}	&	no\\
(our code) &	 0.2245(7)&&
\\\hline
Flavianet'2010~\cite{Antonelli:2010yf} &	0.2253(9) &	 2010~\cite{Antonelli:2010yf}	&	2010~\cite{Antonelli:2010yf}	&	no\\
(our code) &	 0.2254(9)&&
\\\hline
\end{tabular}
\end{center}
\label{tab:comparison}
\end{table}

As a consistency check, we have used our fitting code with the inputs of a few well-known SM analyses and, as shown in Table~\ref{tab:comparison}, we reproduce 
very well their corresponding results. The only difference between our SM analysis and that of Ref.~\cite{Moulson:2014cra} is the inclusion of
the additional inputs discussed above, i.e. $K_{\mu2(\gamma)}$ and the CTT. Thus, the comparison with it shows clearly that their numerical
impact on the $V_{ud}$ and $V_{us}$ values is very small. This is due to the not-precisely-enough values of $f_{K^\pm}$, $\log C_{\rm expt}$ and $\Delta_{\rm CT}$.
In fact, with the lattice values used in this work we find $f_{K^\pm}/f_{\pi^\pm} \times 1/f_+(0) = 1.234(7)$.
Let us stress that the use of $K_{\mu2(\gamma)}$ and CTT was, however, critical in the general NP analysis presented in the previous section.

\section{SMEFT and complementarity with collider searches}
\label{sec:SMEFT}

In order to connect the experimental bounds on the WC obtained at low-energies with those generated by NP models at a high-energy scale, one needs to
take into account the running and mixing under radiative corrections of the corresponding operators. 

At the same time, such scale evolution allows one to make contact with the high-energy SMEFT. Employing this EFT as in intermediate step before
connecting to specific models is convenient for several reasons. First, it is constrained by the more restrictive EW gauge symmetry group which 
leads to model-independent relations among the WC that are not present in the low-energy analysis (\textit{viz.} Eq.~(\ref{eq:RHCuniversal}))~\cite{Cirigliano:2009wk,Alonso:2014csa}.
In addition, and due again to $SU(2)\times U(1)$, the SMEFT Wilson Coefficients enter not only in the processes studied in this work but also in other low-energy 
charged-current or neutral current processes involving first- and second- generation fermions, so that an interesting degree of 
complementarity is expected with charm-hadron (semi)leptonic decays, rare kaon decays, etc. 

Last but not least, the SMEFT makes possible to study model-independently the interplay between the low-energy measurements discussed in this work and NP searches at colliders. 
Although such studies are clearly beyond the scope of this work, in this section we show the potential of this approach through some illustrative and simple examples.

\subsection{RGE running and matching to the SMEFT}

As explained in Sec.~\ref{sec:EFT}, (pseudo)scalar and tensor WC run under QCD and, moreover, they mix through EW interactions. 
We take both effects into account integrating the coupled differential renormalization group equations:
\begin{align}
\frac{d\,\vec\epsilon(\mu)}{d\log\mu}= \left( \frac{\alpha(\mu)}{2\pi}\gamma_{\rm ew} + \frac{\alpha_s(\mu)}{2\pi}\gamma_s \right)\,\vec\epsilon(\mu),
\end{align}
where, once again, $\vec\epsilon(\mu)=(\eS^{D\ell}(\mu),\,\eP^{D\ell}(\mu),\,\eT^{D\ell}(\mu))$. Evolving from the low-energy scale $\mu=2$ GeV to a typical LHC scale such as $\mu=1$ TeV, we find
\bea
\left(
\begin{array}{c}
\epsilon_S^{D\ell}\\
\epsilon_P^{D\ell} \\
\epsilon_T^{D\ell} \\
\end{array}
\right)_{\mbox{($\mu=1$~TeV)}}
= 
\left(
\begin{array}{ccc}
 0.51 & -0.0014 & 0.35 \\
 -0.0014 & 0.51 & 0.35 \\
 0.0031 & 0.0031 & 1.08 \\
\end{array}
\right) 
\left(
\begin{array}{c}
\epsilon_S^{D\ell}\\
\epsilon_P^{D\ell} \\
\epsilon_T^{D\ell} \\
\end{array}
\right)_{\mbox{($\mu=2$~GeV)}}~,
\label{eq:RGE}
\eea
where one can see an important mixing between tensor and (pseudo)scalar, which is simply the result of having large coefficients in 
the corresponding entries of the electroweak anomalous dimension matrix of Eq.~\eqref{eq:EWanomalousdim}.

These results can be trivially used to run the bounds to the high scale:
\bea
\left(
\begin{array}{c}
 \epsilon_S^{de} \\
 \epsilon_P^{de} \\
 \epsilon_T^{de} \\
 \epsilon_S^{se} \\
\epsilon_P^{se} \\
 \epsilon_T^{se} \\
 \epsilon_S^{s\mu} \\
 \epsilon_P^{s\mu}  \\
 \epsilon_T^{d\mu} \\
\end{array}
\right) 
=
\left(
\begin{array}{c}
 7.4\pm 7.1 \\
 0.3\pm 2.8 \\
 1.1\pm 8.7 \\
 2.3\pm 6.5 \\
 3.1\pm 6.2 \\
 1.0\pm 1.9 \\
 0.0\pm 1.8 \\
 -0.2\pm 2.9 \\
 0.6\pm 5.6 \\
\end{array}
\right)
\times 10^{\wedge}
\left(
\begin{array}{c}
 -4 \\
 -4 \\
 -4 \\
 -3 \\
 -3 \\
 -2 \\
 -3 \\
 -3 \\
 -3 \\
\end{array}
\right)~,
\eea
in the $\overline{MS}$ scheme at $\mu=1$ TeV. The corresponding correlation matrix is given in Appendix~\ref{app:WC1TeV}. The mixing 
between operators produces larger diagonal errors for the pseudoscalar WC and induce very large non-diagonal entries. As explained 
before such large correlations reflect the fact that certain WC combination are much more constrained than the individual couplings.

With the values of the $\epsilon_i$ expressed at the high-energy scale, one can now translate them into determinations of the
WC of the SMEFT, that we will denote as $\alpha_i$. The (tree-level) matching equations between the low-energy EFT and the SMEFT are~\cite{Cirigliano:2009wk}:
\bea
\frac{\delta G_F}{G_F} &=& 2~[\hat{\alpha}_{\varphi l}^{(3)}]_{11+22} - [\hat{\alpha}_{ll}^{(1)}]_{1221} - 2 [\hat{\alpha}_{ll}^{(3)}]_{1122-\frac{1}{2}(1221)}~,\nonumber\\
V_{1j} \cdot
\epsilon_L^{j\ell}
&=& 2 \, V_{1j}  \,  \left[\hat{\alpha}_{\varphi l}^{(3)}\right]_{\ell\ell}   +   2 \,\left[V\hat{\alpha}_{\varphi q}^{(3)}\right]_{1j} -    2\, \left[V\hat{\alpha}_{l q}^{(3)}\right]_{\ell\ell 1j}, \nonumber\\
V_{1j} \cdot  \epsilon_R^{j} &=& - \left[\hat{\alpha}_{\varphi \varphi}\right]_{1j}, \nonumber\\
V_{1j} \cdot \epsilon_{s_L}^{j\ell} &=& - \left[\hat{\alpha}_{l q}\right]_{\ell\ell j1}^*, \nonumber\\
V_{1j} \cdot \epsilon_{s_R}^{j\ell} &=& -  \left[V\hat{\alpha}^\dagger_{qde}\right]_{\ell\ell 1j},\nonumber  \\
V_{1j} \cdot  \epsilon_T^{j\ell} &=& - \left[\hat{\alpha}^t_{l q} \right]_{\ell\ell j1}^* ~,
\label{eq:matchingeqs}
\eea
where we 
labeled the quark generations with numbers, and introduced, for simplicity, 
$\epsilon_{s_{L/R}}^{j\ell}=(\epsilon_S^{j\ell}\pm\epsilon_P^{j\ell})/2$. 
In Eqs.~(\ref{eq:matchingeqs}) the repeated indices $j,\ell$ are not summed over, while the index $m$ is. Finally let us also 
notice that $2\hat{\alpha}_i = \alpha_i ~v^2/\Lambda^2$, with $v=(\sqrt{2}G_F)^{-1/2}\simeq 246$ GeV.

The matching equation for $\epsilon_R^D$ shows clearly that this WC is lepton independent, \textit{c.f.} Eq.~(\ref{eq:RHCuniversal}), at this order in the SMEFT expansion, since the corresponding operator is ${\cal O}_{\varphi\varphi} = i (\varphi^T\epsilon D_\mu \varphi)(\bar{u}\gamma^\mu d)$ + h.c.

Concerning the complementarity with collider searches, it is useful to notice the difference between chirality-conserving and -violating operators. On one hand, we have the SMEFT Wilson Coefficients contributing to $\delta G_F$ and $\eL^{j\ell}$, and probed in our fit through $\Delta_{\rm CKM},\Delta_{LP}^d$ and $\Delta_L^s$, which conserve chirality. Their interference with the SM is not suppressed in collider observables and an interesting interplay with LEP~\cite{Efrati:2015eaa,Falkowski:2015krw} and LHC~\cite{Bhattacharya:2011qm,Cirigliano:2012ab,deBlas:2013qqa} searches is expected.

On the other hand, the SMEFT Wilson Coefficients contributing to $\eR^d,\eS^{s\ell},\eP^{s\ell}$ and $\eT^{s\ell}$ are chirality-flipping, and thus they are not accessible by LEP searches at order $v^2/\Lambda^2$. In the case of (pseudo)scalar and tensor operators, the LHC can still provide interesting limits, thanks to the ${\cal O}(s^2/v^4)$ enhancement of the quadratic term (due to their contact-interaction nature)~\cite{Bhattacharya:2011qm,Cirigliano:2012ab}, as we will discuss in Section~\ref{sec:LHC}. However, for $\eR^d$ this is not the case, as it is generated by non-standard $W$ couplings to right-handed quarks. The (semi)leptonic decays studied in this paper provide clearly a unique probe for these operators without competitors in the collider frontier.

Finally, in the matching equations of Eq.~\eqref{eq:matchingeqs} we used the basis of operators employed in Ref.~\cite{Cirigliano:2009wk}, which was a modified version of the seminal Buchmuller-Wyler basis~\cite{Buchmuller:1985jz} with the relevant redundancies (and the addition of one missing operator) properly taken care of.\footnote{
One redundancy was not eliminated on purpose (either $O_{ll}^{(3)}$ or  $O_{ll}^{(1)}$), so that each operator is proportional to the unit matrix in the $U(3)^5$-symmetric case~\cite{Cirigliano:2009wk}. The Warsaw basis~\cite{Grzadkowski:2010es} chooses instead to keep only the operator $O_{ll}^{(1)}$, which in the $U(3)^5$-symmetric limit has two independent flavor contractions.} Additional redundancies in other sectors of the Buchmuller-Wyler basis (not relevant for semileptonic quark decays) were later identified in Ref.~\cite{Grzadkowski:2010es}, where the first minimal and complete SMEFT basis was derived. In the sector relevant for our work, this so-called Warsaw basis is in fact very similar to the one of Ref.~\cite{Cirigliano:2009wk} used in this work, up to some numerical factors and conventions. In particular, in these bases the equations of motions were not used to remove any operator containing fermions in favor of purely fermionic operators, which would introduced some complications in our flavor general analysis.

\subsection{MFV limit}
Once again it is interesting to have a look at the $U(3)^5$-symmetric limit, which is in practice equivalent to the MFV case. In that
limit we recover the result obtained in Ref.~\cite{Cirigliano:2009wk}, namely
\bea
\Delta_{\rm CKM} = 2~\frac{v^2}{\Lambda^2}~\left( - \alpha_{\varphi l}^{(3)} + \alpha_{\varphi q}^{(3)}- \alpha_{\ell q}^{(3)} + \alpha_{ll}^{(3)}        \right)~.
\eea
The bound on this combination of WC from (semi)leptonic hadron decays, $\Delta_{\rm CKM} = -(4.6\pm 5.2)\times 10^{-4}$, \textit{c.f.} 
Eq.~\eqref{eq:deltackm}, corresponds to an effective scale $\Lambda>10$ TeV (90\%CL). As shown in Ref.~\cite{Cirigliano:2009wk}, such bound is much stronger than the
limit obtained from the combined analysis of LEP and other EW precision observables. Thus, it is an important input for global EFT fits performed in 
this limit~\cite{Han:2004az,Berthier:2015gja}.\footnote{
Ref.~\cite{Berthier:2015gja} does not work in the $U(3)^5$-symmetric case, but in a more restrictive scenario, since the two independent contractions of flavor indices allowed by the $U(3)^5$ symmetry for the operator $Q_{\ell\ell}$ (corresponding to $\alpha_{ll}^{(1)}$ and $\alpha_{ll}^{(3)}$ in our basis) are controlled by one single coefficients $C_{\ell\ell}$ in that work.} Finally let us notice that this is even more the case if a non-linear EFT framework is used, since more operators have to be considered~\cite{Brivio:2016fzo}.

\subsection{Bounds on scalar and tensor interactions}
\label{sec:LHC}

If the new particles are too heavy to be produced on-shell at the LHC we can connect collider searches with 
low-energy processes in an elegant model-independent way using the SMEFT~\cite{Bhattacharya:2011qm,Cirigliano:2012ab}. Since we
are interested in non-standard effects in semileptonic $D$-quark decays, $D\to u \ell\bar{\nu}$, the natural channel
to study at the LHC is $pp\to \ell+ {\rm MET}+X$, since this process is sensitive at tree level to non-standard 
$\bar{u}D\to \ell\bar{\nu}$ partonic interactions. The comparative analysis between the bounds from nuclear $\beta$ decays and the 
LHC for WC involving the $d$-quark was performed in Refs.~\cite{Bhattacharya:2011qm,Cirigliano:2012ab}. 
The study was extended to the hyperon $\beta$ decays for the operators involving the $s$-quark in Ref.~\cite{Chang:2014iba}, and we
extend it further here by comparing the LHC bounds and those obtained for $\epsilon_{S,T}^{s\ell}$ from $K_{\ell 3}$.

First we briefly explain how the LHC bounds were obtained in Refs.~\cite{Bhattacharya:2011qm,Cirigliano:2012ab,Chang:2014iba}. 
By using the matching relations in Eqs.~(\ref{eq:matchingeqs}), one can express the cross-section $\sigma(pp\to \ell+{\rm MET}+X)$
as is modified by non-standard $\bar{u}s\to \ell\bar{\nu}$ partonic interactions:
\bea
\label{eq:sigmamt}
\sigma(m_T \!\!>\! \overline{m}_T) &=&
\sigma_W 
+ \sigma_S|\eS^{s\ell}|^2 +  \sigma_T |\eT^{s\ell}|^2 ~,
\eea  
where $\sigma_W (\overline{m}_T)$ represents the SM contribution and $\sigma_{S,T}(\overline{m}_T)$ 
are new functions, with transverse mass higher than $\overline{m}_T$, which explicit form can be found in Ref.~\cite{Cirigliano:2012ab}.
The crucial feature is that they are several orders of magnitudes larger than the SM contribution,
what compensates for the smallness of the NP couplings and makes possible to put significant bounds on them from these searches.
Thus, comparing the observed events above $\overline{m}_T$ with the SM expectation we can set bounds on $\epsilon_{S,T}^{s\ell}$. 

Some caveats are in order. First, it is important to note that the dependence of the cross section~(\ref{eq:sigmamt}) on the WC is quadratic. 
We assume that contributions from SMEFT dimension-8 operators can be neglected, which is expected to happen for a broad class of NP models 
(see e.g. Ref.~\cite{Contino:2016jqw}). And secondly, this cross section is sensitive to a plethora of other dimension-6 effective operators, 
some of them interfering with the SM, which make possible the appearance of flat directions that we neglect here. It is worth stressing that 
these assumptions were not necessary in the low-energy fit.

\begin{figure}[b!]
\centerline{
\includegraphics[width=0.42\textwidth]{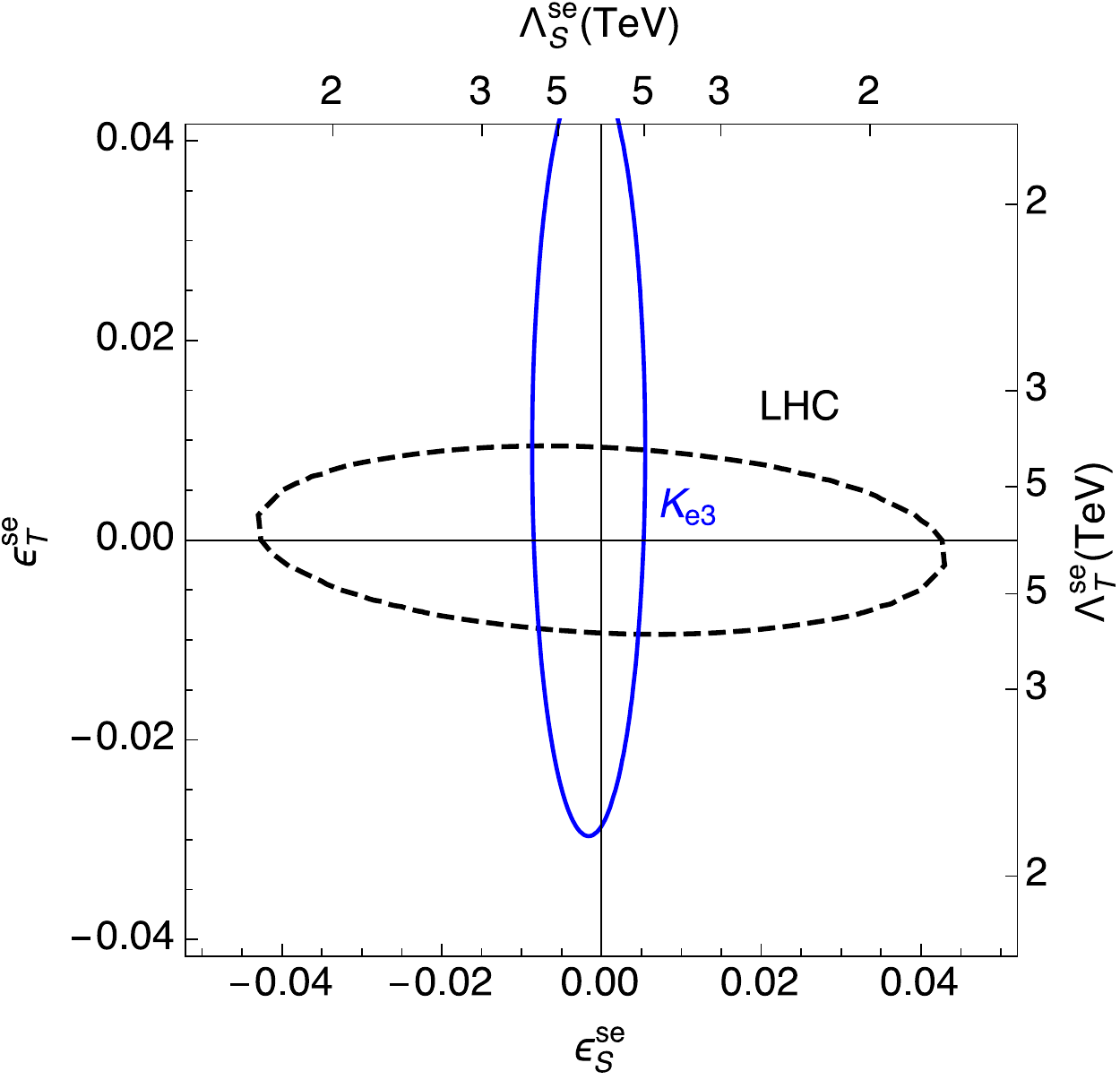}
\hspace{1cm}
\includegraphics[width=0.42\textwidth]{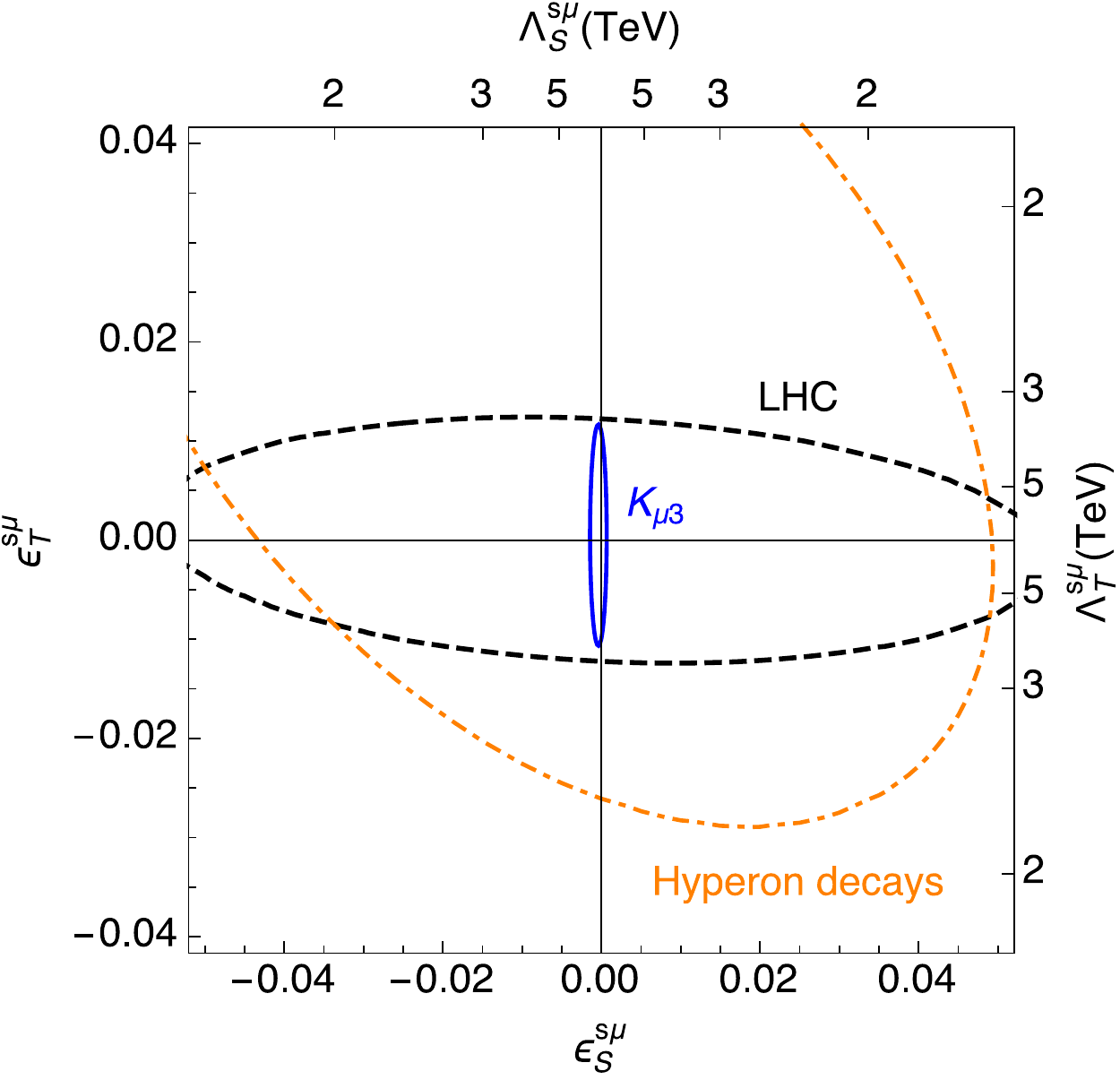}}
\vspace{0.5cm}
\centerline{
\includegraphics[width=0.42\textwidth]{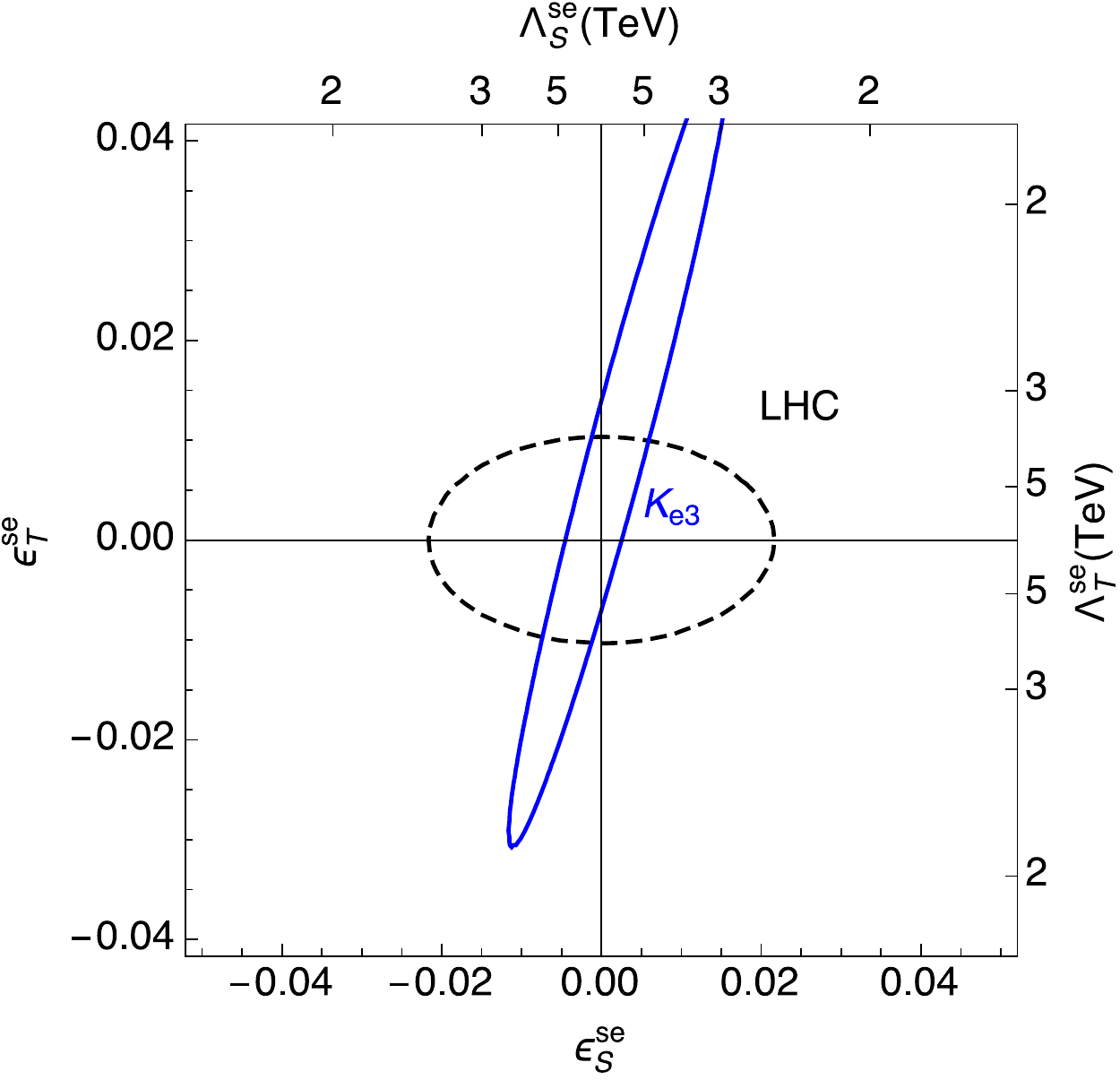}
\hspace{1cm}
\includegraphics[width=0.42\textwidth]{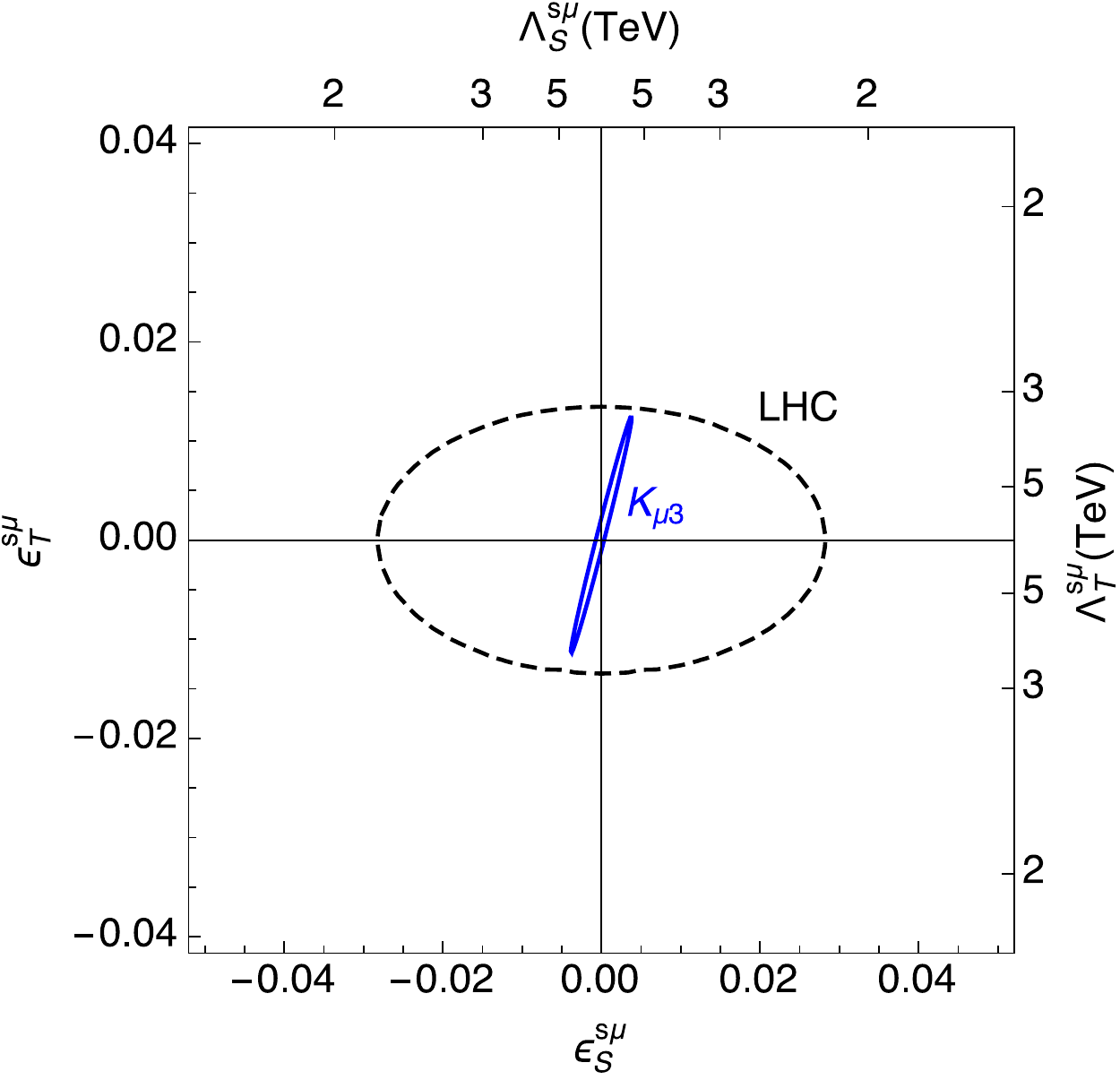}}
\caption{90\% CL constraints on $\epsilon_{S,T}^{s\ell}$ from our global fit (blue solid ellipse), from the analysis 
of $pp\to \ell^\pm+{\rm MET}+X$ CMS data (black dashed ellipse) and from semileptonic hyperon decays (orange dot-dashed lines)~\cite{Chang:2014iba}. 
The left (right) panel corresponds to the electronic (muonic) couplings, whereas the upper (lower) panel show the results at $\mu=2$ GeV ($1$ TeV). Effective scales are defined by $\Lambda_i \approx (V_{us} \epsilon_i)^{-1/2} v$,  
see Eqs.~(\ref{eq:matchingeqs}).
\label{fig:CPST}}   
\end{figure}

Using 20 fb$^{-1}$ of data recorded at $\sqrt{s}=$ 8 TeV by the CMS collaboration in the electron channel $pp\to e^\pm+{\rm MET}+X$\cite{Khachatryan:2014tva}, 
and choosing $\overline{m}_T=1.5$~TeV, the 90\% C.L. limit shown in Fig.~\ref{fig:CPST} (left panel) was obtained in Ref.~\cite{Chang:2014iba}. 
In particular, one event is found with a transverse mass above $\overline{m}_T=1.5$~TeV in the 20 fb$^{-1}$ dataset recorded at $\sqrt{s}=$ 8 TeV by the 
CMS collaboration~\cite{Khachatryan:2014tva}, in good agreement with the SM background of $2.02\pm 0.26$ events. We repeat the same analysis here for the 
muonic channel  $pp\to \mu^\pm+{\rm MET}+X$\cite{Khachatryan:2014tva}, where good agreement is also observed between data (3 events above $\overline{m}_T=1.5$~TeV)
and SM ($2.35\pm 0.70$ events), obtaining the bound shown in Fig.~\ref{fig:CPST} (right panel). The terms $\sigma_{S,T}$ were calculated using the MSTW2008 PDF
sets evaluated at $Q^2=1$ TeV$^2$~\cite{Martin:2009iq}. Further details can be found in Ref.~\cite{Cirigliano:2012ab}. The running of the limits from 1 TeV to 2 GeV
is performed using the QCD+EW RGE, \textit{c.f.} Eq.~\eqref{eq:RGE}.

The figure shows also the limits obtained in Ref.~\cite{Chang:2014iba} from the $R_{B_1B_2}$ ratio in Eq.~\eqref{eq:SHDNP} and the existing data in several 
semileptonic hyperon decays, and the limits obtained in this work from the global analysis of $D\to u \ell\bar{\nu}$ processes.

Fig.~\ref{fig:CPST} illustrates the interesting complementarity between low-energy experiments and the LHC searches. While the LHC cross section is almost 
equally sensitive to both electron and muon couplings, $K_{\ell3(\gamma)}$ is much more sensitive to the muon one, where the interference with the SM is 
the dominant contribution. The figure illustrates quite clearly that semileptonic kaon decays are exploring new regions in the NP parameter space unaccessible
for the LHC and corresponding to 1-10 TeV effective scales. Finally, the lower panel of Fig.~\ref{fig:CPST} shows the same limits, this time at $\mu=1$ TeV, which might be more interesting
from a model-building perspective.

Needless to say, this interplay becomes much more interesting if a discrepancy with the SM is found. This was illustrated in Ref.~\cite{Bhattacharya:2011qm} 
assuming the presence of a scalar resonance at the LHC that, in our case, could be detected as a nonzero $\epsilon_S^{s\ell}$ in $K_{\ell 3}$
(or a non-zero pseudoscalar/tensor coupling due to the RGE mixing).

\section{Conclusions}
\label{sec:Conclusions}

In this paper we introduce a global model-independent analysis of NP in the $D\to u\ell\nu$ transitions ($D=d,s$; $\ell=e,\,\mu$)
in the context of the SMEFT. We do not assume any flavor symmetry and we keep all possible NP operators at the same time.  
Special attention is paid to the (semi)leptonic kaon decays where such a comprehensive and systematic 
analysis of NP was lacking, and we study the complementarity with pion decays and nuclear, neutron and hyperon $\beta$ decays. 
The latter become necessary since one can not discriminate among all different possible NP effects using only pion and kaon decay observables.
This is not only relevant for the determination of $|\tilde V_{ud}^e|$ but also for singling out the effects of NP contributions from right-handed currents. 
In this sense, future analyses would greatly benefit from a better understanding of neutron and hyperon properties such as $g_A$ and $g_1$. 

Besides providing a road map for future tests of the SM using all these processes, we provide numerical results of a fit using current experimental 
data and lattice QCD results. Our analysis includes the MFV and the SM limit as a specific case. In fact our output are not only the bounds on the 
various WC, but also the $V_{us}$ and $V_{ud}$ elements, and includes various QCD form factors parameters. In the SM 
limit we recover the most precise determinations of them, with small improvements due to the inclusion of the individual rate of $K_{\mu2}$ as 
a separate input and the Callan-Treiman theorem. 

We find that these decays are sensitive to NP with typical scales of several TeV,
especially in the case of a pseudoscalar contribution to $P^\pm_{e2}$ ($P=K,\,\pi$), which is ruled out up to scales as high as $\mathcal{O}(100)$ TeV.
To make this connection to the high-energy scale more explicit, we properly accounted for the operator running and mixing under the QCD and EW interactions, we 
expressed the low-energy bounds in the context of the SMEFT and provided their value at 1 TeV. Our results can then be matched straightforwardly to 
any specific NP model. Very large correlations appear between different WC, which should have non-trivial implications for the chosen model. Finally, 
as illustrated with a few simple cases in this work, the matching with the SMEFT opens the possibility for powerful synergies between these decays 
and searches of NP at the LHC.

\section*{Acknowledgements}
The authors thank L.~Alvarez-Ruso, V.~Cirigliano, J.~C.~Hardy, V.~Lubicz, M.~Moulson, A.~Pich, J.~Portol\'es, I.~C.~Towner, M.~Vicente Vacas, F.~Wauters, and O.~Yushchenko for useful discussions and correspondence. We would also like to acknowledge the Institute for Nuclear Theory (INT) and the Mainz 
Institute for Theoretical Physics (MITP) for enabling us to complete a significant portion of this work. 
M.G.-A. is grateful to the LABEX Lyon Institute of Origins (ANR-10-LABX-0066) of the Universit\'e de Lyon 
for its financial support within the program ANR-11-IDEX-0007 of the French government. JMC's work is funded 
by the People Programme (Marie Curie Actions) of the European Union's Seventh Framework Programme (FP7/2007-2013)
under REA grant agreement n PIOF-GA-2012-330458.

\newpage

\appendix

\section{WC bounds at $\mu=1$ TeV}
\label{app:WC1TeV}
 
\bea
\left(
\begin{array}{c}
 \tilde{V}_{ud}^{e} \\
 \Delta_{\rm CKM}\\
 \Delta_L^s \\
 \eL^{de}-\eL^{d\mu} +  0.22\eS^{d\mu}+47 \eP^{d\mu}-15 \eT^{d\mu}\\
 \epsilon_R^d\\
 \epsilon_R^s \\
 \epsilon_S^{de} \\
 \epsilon_P^{de} \\
 \epsilon_T^{de} \\
 \epsilon_S^{se} \\
\epsilon_P^{se} \\
 \epsilon_T^{se} \\
 \epsilon_S^{s\mu} \\
 \epsilon_P^{s\mu}  \\
 \epsilon_T^{s\mu} \\
\end{array}
\right) 
=
\left(
\begin{array}{c}
 0.97451\pm 0.00038 \\
 -1.2\pm 8.4 \\
 1.0\pm 2.5 \\
 1.9\pm 3.8 \\
 -1.3\pm 1.7 \\
 0.1\pm 5.0 \\
 7.4\pm 7.1 \\
 0.3\pm 2.8 \\
 1.1\pm 8.7 \\
 2.3\pm 6.5 \\
 3.1\pm 6.2 \\
 1.0\pm 1.9 \\
 0.0\pm 1.8 \\
 -0.2\pm 2.9 \\
 0.6\pm 5.6 \\
\end{array}
\right)\times 10^{\wedge}\left(
\begin{array}{c}
 0 \\
 -4 \\
 -3 \\
 -2 \\
 -2 \\
 -2 \\
 -4 \\
 -4 \\
 -4 \\
 -3 \\
 -3 \\
 -2 \\
 -3 \\
 -3 \\
 -3 \\
\end{array}
\right),
\eea
in the $\overline{MS}$ scheme at $\mu=1$ TeV. Let us remind the reader that $\Delta_L^s = \epsilon_{L}^{s\mu}-\epsilon_{L}^{se} $.
The correlation matrix is given by:
\bea
\rho=
\left(
\scriptsize{
\begin{array}{ccccccccccccccc}
 1. & 0.88 & 0. & 0.01 & 0. & 0. & 0.76 & -0.01 & 0. & 0. & 0. & 0. & 0. & 0. & 0. \\
 - & 1. & -0.07 & 0.01 & 0. & 0. & 0.67 & 0. & 0. & 0. & 0. & 0. & -0.02 & 0.01 & 0. \\
 - & - & 1. & 0. & 0. & 0. & 0. & 0. & 0. & 0. & 0. & 0. & 0.45 & 0.3 & 0.46 \\
 - & - & - & 1. & -0.87 & 0. & 0.01 & 0.01 & 0. & 0. & 0. & 0. & 0.01 & 0.07 & 0. \\
 - & - & - & - & 1. & 0. & 0. & -0.01 & 0. & 0. & 0. & 0. & 0. & 0. & 0. \\
 - & - & - & - & - & 1. & 0. & 0. & 0. & 0. & 0. & 0. & 0. & -0.75 & 0. \\
 - & - & - & - & - & - & 1. & 0.38 & 0.39 & 0. & 0. & 0. & 0. & 0. & 0. \\
 - & - & - & - & - & - & - & 1. & 0.9998 & 0. & 0. & 0. & 0. & 0. & 0. \\
 - & - & - & - & - & - & - & - & 1. & 0. & 0. & 0. & 0. & 0. & 0. \\
 - & - & - & - & - & - & - & - & - & 1. & 0.96 & 0.97 & 0. & 0. & 0. \\
 - & - & - & - & - & - & - & - & - & - & 1. & 0.999998 & 0. & 0. & 0. \\
 - & - & - & - & - & - & - & - & - & - & - & 1. & 0. & 0. & 0. \\
 - & - & - & - & - & - & - & - & - & - & - & - & 1. & 0.63 & 0.99 \\
 - & - & - & - & - & - & - & - & - & - & - & - & - & 1. & 0.64 \\
 - & - & - & - & - & - & - & - & - & - & - & - & - & - & 1. \\
\end{array}
}
\right)~.
\eea

\bibliographystyle{JHEP}
\bibliography{draftv2arxiv}

\end{document}